\documentclass[a4paper,11pt]{article}
\pdfoutput=1 

\usepackage{amsfonts}
\usepackage{amssymb, amscd}
\usepackage{graphicx}
\usepackage{mathrsfs}
\usepackage{slashed}
\usepackage{pdfpages}
\usepackage{braket}
\usepackage{sidecap}
\usepackage{arydshln}
\usepackage[font=small,labelsep=none]{caption}
\usepackage{mathtools}
\usepackage{hyperref}
\usepackage{tikz}
\usetikzlibrary{arrows,decorations.markings}
\definecolor{myblue}{RGB}{174, 198, 219}
\definecolor{myred}{RGB}{157,31,68}
\usepackage{subcaption}
\definecolor{ceruleanblue}{rgb}{0.0, 0.2, 0.6}
\usepackage[left=2cm,right=2cm,top=3.5cm,bottom=3.5cm]{geometry}
\newcommand{\be}{\begin{equation}}      
\newcommand{\ee}{\end{equation}}

\numberwithin{equation}{section}

\hypersetup{
    colorlinks=true,
    linkcolor=ceruleanblue,
    filecolor=ceruleanblue,      
    urlcolor=ceruleanblue,
    citecolor=ceruleanblue,
}

\def\MP{M_{\rm Pl}}

\newcommand{\betas}{\beta}%
\newcommand{\vect}[1]{\boldsymbol{#1}}

\begin{document}

\begin{center}
\LARGE{\bf Resonant Decay of Gravitational Waves\\ into Dark Energy}
\\[1cm] 

\large{Paolo Creminelli$^{\,\rm a, \rm b}$, Giovanni Tambalo$^{\,{\rm c},{\rm d}}$, Filippo Vernizzi$^{\, \rm e}$ and Vicharit Yingcharoenrat$^{{\,\rm c }, {\rm d}}$}
\\[0.5cm]

\small{
\textit{$^{\rm a}$
ICTP, International Centre for Theoretical Physics\\ Strada Costiera 11, 34151, Trieste, Italy}}
\vspace{.2cm}

\small{
\textit{$^{\rm b}$
IFPU - Institute for Fundamental Physics of the Universe,\\ Via Beirut 2, 34014, Trieste, Italy }}
\vspace{.2cm}

\small{
\textit{$^{\rm c}$ SISSA, via Bonomea 265, 34136, Trieste, Italy}}
\vspace{.2cm}

\small{
\textit{$^{\rm d}$ INFN, National Institute for Nuclear Physics \\  Via Valerio 2, 34127 Trieste, Italy}}
\vspace{.2cm}

\small{
\textit{$^{\rm e}$ Institut de physique th\' eorique, Universit\'e  Paris Saclay, CEA, CNRS \\ [0.05cm]
 91191 Gif-sur-Yvette, France}}
\vspace{.2cm}

\end{center}

\vspace{0.3cm} 

\begin{abstract}\normalsize
We study the decay of gravitational waves into dark energy fluctuations $\pi$, taking into account the large occupation numbers. 
We describe dark energy using the effective field theory approach, in the context of generalized scalar-tensor theories. When the  $m_3^3$ (cubic Horndeski)
and $\tilde m_4^2$ (beyond Horndeski) operators are present, 
the gravitational wave acts as a classical background for $\pi$ and modifies its dynamics. In particular, $\pi$ fluctuations are described by a Mathieu equation and feature instability bands that grow exponentially. 
Focusing on the regime of small gravitational-wave amplitude, corresponding to narrow resonance, we calculate analytically the produced $\pi$, its energy and the change of the gravitational-wave signal. 
The resonance is affected by $\pi$ self-interactions in a way that we cannot describe analytically. 
This effect is very relevant for the operator $m_3^3$ and it limits the instability. In the case of the $\tilde m_4^2$ operator self-interactions can be neglected, at least in some regimes. The modification of the gravitational-wave signal is observable for \textcolor{black}{$3 \times 10^{-20} \lesssim \alpha_{\rm H} \lesssim 10^{-17}$} with a LIGO/Virgo-like interferometer and for $10^{-16} \lesssim \alpha_{\rm H} \lesssim 10^{-10}$ with a LISA-like one. 

\end{abstract}

\vspace{0.3cm} 

\newpage 
{
  \hypersetup{linkcolor=black}
  \tableofcontents
}

\vspace{0.3cm}

\flushbottom

\vspace{0.3cm}

\section{Introduction}

In a recent article \cite{Creminelli:2018xsv}, some of us pointed out that in modified-gravity theories gravitational waves (GWs) can decay into dark energy fluctuations, as a consequence of the spontaneous breaking of
Lorentz invariance. 
We focused on dark energy and modified gravity   based on a scalar degree of freedom, whose time-dependent background induces a preferred time-foliation of the  FRW metric. The natural way to describe this setting is the effective field theory of dark energy (EFT of DE) \cite{Creminelli:2006xe,Cheung:2007st,Creminelli:2008wc,Gubitosi:2012hu,Gleyzes:2013ooa}, which we review below.  Moreover, since the relative speed between GWs and light is now  constrained with $10^{-15}$ accuracy \cite{TheLIGOScientific:2017qsa},  we  considered only models where  gravitons travel at the same speed as light \cite{Creminelli:2017sry,Sakstein:2017xjx,Ezquiaga:2017ekz,Baker:2017hug}. (One possible caveat about these constraints is that they do not apply if the theory breaks down at scales parametrically lower than the frequency of the observed GWs \cite{deRham:2018red}. In this case the propagation of GWs, as well as many accurate tests of gravity, can only be described if one knows the UV completion of the EFT of DE. )

In particular, we showed that some of the  operators of the EFT of DE
display a cubic $\gamma  \pi \pi$ interaction, where $\gamma$ and $\pi$ respectively denote the graviton and the scalar-field fluctuation, that can mediate the decay. Of course, this can happen only if both energy and momentum can remain conserved during the process, which is when scalar  fluctuations propagate subluminally.
The same vertex is also responsible for an anomalous GW dispersion, for speeds of scalar propagation different from that of light. 
Depending on the energy scale suppressing this interaction, these two effects can be important at frequencies observed by LIGO/Virgo and can constrain these theories.

The bound derived in \cite{Creminelli:2018xsv} is based on a perturbative calculation, in which individual gravitons are assumed to decay independently of each other. But a classical GW is a collection of many particles with very large occupation number and particle production must be treated as a collective process in which
many gravitons decay simultaneously.  The classical GW acts as a background for the propagation of scalar fluctuations. This paper studies this process in the limit where the GW background acts as a small periodic perturbation (narrow resonance). Larger amplitudes of the GW can induce tachyon or ghost instabilities: we will study this possibility in a forthcoming publication \cite{Creminelli:2019new}.

To review the EFT approach, 
we define the necessary geometrical quantities. 
We assume the flat FRW metric to be given by $\text d s^2 = - \text d t^2 + a^2(t) \text d \vect x^2 $, so that $\delta g^{00}\equiv 1+g^{00}$ is the perturbation of $g^{00}$ around the background solution. Moreover, we define by $\delta K^\mu_{\ \nu} \equiv K^\mu_{\ \nu}-H\delta^\mu_{\ \nu}$, where $H \equiv \dot a/a$ is the Hubble rate,  the perturbation of the extrinsic curvature of the equal-time hypersurfaces and  by $\delta K$ its trace. Finally, we will need the 3d Ricci scalar of these hypersurfaces, ${}^{(3)}\! R$. 

As mentioned above, we focus on theories where gravitons travel luminally and, for later convenience, we split the EFT of DE action in the sum of three actions \cite{Creminelli:2017sry}, 
\be
\label{total_action}
S = S_0 + S_{m_3} + S_{\tilde m_4} \;,
\ee
where
\begin{align}
\label{starting_action}
S_0  & =  \int  \text d^4 x \sqrt{-g}  \bigg[  \frac{\MP^2}{2}  \, {}^{(4)}\!R - \lambda(t)- c(t) g^{00}  +  \frac{m_2^4(t)}{2} (\delta g^{00})^2  \bigg] \;, \\
S_{m_3} & = - \int  \text d^4 x  \sqrt{-g} \, \frac{m_3^3(t)}{2}\delta K\delta g^{00} \;, \label{eq:m3term}\\
S_{\tilde m_4} & =\int  \text d^4 x   \sqrt{-g} \, \frac{\tilde{m}_4^2 (t)}{2} \delta g^{00} \left({}^{(3)}R + \delta K_\mu^\nu \delta K_\nu ^\mu  - \delta K^2 \right) \;. \label{eq:bh}
\end{align}
The first action, $S_0$, contains the Einstein-Hilbert term and  the minimal scalar field Lagrangian, which describe the dynamics of the background. 
Notice that we have removed any time-dependence in front of the Einstein-Hilbert term by a conformal transformation, which leaves the graviton speed unaffected.
Indeed, the first two functions of time, $c(t)$ and $\lambda(t)$,\footnote{We use the notation $\lambda(t)$, and not $\Lambda(t)$ as usual in the  literature, to avoid confusion with the energy scale $\Lambda$ suppressing the cubic vertices studied in this article.} can be determined by the background evolution in terms of the Hubble expansion and matter quantities {(see e.g.~Sec. 5 of \cite{Gubitosi:2012hu} for their explicit expressions in this frame).}  The operator $m_2^4$ does not change the background but affects the speed of propagation of scalar fluctuations, $c_s^2$. Its typical value is ${\cal O}(\MP^2 H^2)$, see e.g.~\cite{Gubitosi:2012hu} for details. This action was studied in details in \cite{Creminelli:2008wc} and
in the covariant language it describes quintessence \cite{Caldwell:1997ii} or, more generally, a  dark energy with scalar field  Lagrangian $P(\phi, X)$ \cite{ArmendarizPicon:2000dh}, where $X \equiv g^{\mu \nu} \partial_\mu \phi \partial_\nu \phi $.

 The operator in the second action, $S_{m_3}$,  introduces a kinetic mixing between the scalar field and gravity \cite{Creminelli:2006xe,Creminelli:2008wc} (sometimes called kinetic gravity braiding \cite{Deffayet:2010qz,Kobayashi:2010cm}). In the covariant language it corresponds to the cubic Horndeski Lagrangian \cite{Horndeski:1974wa,Deffayet:2011gz}, of the form  $ Q(\phi, X) \Box \phi$ with $\Box \equiv g^{\mu \nu} \nabla_\mu \nabla_\nu$. In the regime that leads to sizeable modifications of gravity (i.e.~for $m_3^3 \sim \MP^2 H_0$), the operator contained in $S_{m_3}$ displays a $\gamma \pi \pi$ interaction suppressed by an energy scale of order $\Lambda_2 \equiv(\MP H_0)^{1/2} \sim 10^{-3} $eV. This energy scale is much greater than the typical LIGO/Virgo frequency. 
 For this reason, in \cite{Creminelli:2018xsv} the parameter $m_3^3$ remains unconstrained by the graviton decay  computed in perturbation theory. 
Finally, the operator in the third action, $S_{\tilde m_4}$ whose typical size is $\MP H_0$ introduces a kinetic mixing between the scalar field and matter \cite{Gleyzes:2014dya}  (or kinetic matter mixing \cite{DAmico:2016ntq}). In the covariant language it corresponds to the beyond-Horndeski Lagrangian of the Gleyzes-Langlois-Piazza-Vernizzi type \cite{Gleyzes:2014dya,Gleyzes:2014qga}.
This operator displays a $\gamma \pi \pi$ interaction and was constrained in \cite{Creminelli:2018xsv} with the perturbative decay because the vertex is 
 suppressed by an energy scale close to LIGO/Virgo frequencies. 
 
In Sec.~\ref{Sec2}, after expanding the action $S_0+S_{m_3}$ in perturbations in the Newtonian gauge (the same calculation  is repeated in the spatially-flat gauge in App.~\ref{app:spatflat}),  we study the effect of a classical GW background on the $\pi$ dynamics, for the operator $m_3^3$ (Sec.~\ref{subsec2.1}) and $\tilde m_4^2$ (Sec.~\ref{tildem42op}).
The regime of small GWs  can be studied analytically and leads to the so-called {\em narrow} parametric resonance, which is the subject of Sec.~\ref{sec:Resonance}.
There we compute  the  energy density of $\pi$ produced by the parametric instability due to the oscillating GWs  (in Sec.~\ref{subsec:pienergy}) and we re-interpret 
the $\pi$ production in the narrow-resonance regime as an effect of Bose enhancement of the perturbative decay in  App.~\ref{Bose}. The back-reaction on the GW signal is computed in Sec.~\ref{subsec:waveformmod} for a linearly polarized wave, while the case of elliptical polarization is discussed in Sec.~\ref{sec:genpol}. In Sec.~\ref{sec:conservation} we check that energy is conserved in this process, as expected (the details of the calculations are given in App.~\ref{app:energy}). 
 
The treatment in Sec.~\ref{sec:Resonance} neglects scalar-field nonlinearities, which are studied in Sec.~\ref{sec:selfint}.
The operator $m_3^3$ contains cubic self-interactions suppressed by the scale $\Lambda_3 \equiv(\MP H_0^2)^{1/3} \sim 10^{-13}$eV, which is much smaller than the one appearing in the vertex $\gamma \pi \pi$. Thus,  these  become relevant and probably halt the parametric resonance well before the GWs are affected by the back-reaction (Sec.~\ref{sec:nlt}). This makes the results of Sec.~\ref{sec:Resonance} applied to this operator inconclusive.
The situation is different for the operator $\tilde m_4^2$: in this case the scale that suppresses non-linearities is the same that appears in the coupling $\gamma \pi \pi$. The leading non-linearities are quartic in the regime of interest and are suppressed with respect to a naive estimate due to the particular structure of Galileon interactions. At least in some region of parameters non-linearities do not halt the parametric instability due to the oscillating GWs. In Sec.~\ref{sec:mtilde41} we therefore study in which range of parameters one expects a modification of the GW signal.
 Moreover, in Sec.~\ref{sec:mtilde42} we  discuss {\em precursors}, higher harmonics  induced in the GW signal by the produced $\pi$, that enter the observational band earlier than the main signal. 
We conclude discussing the main results of the article and possible future directions in Sec.~\ref{sec:conclusion}.

\section{Graviton-scalar-scalar vertices}
\label{Sec2}

Let us derive the interaction $\gamma\pi\pi$ from the action   \eqref{total_action}, using the Newtonian gauge.  For the time being, we neglect the self-interactions of the $\pi$ field; they will be discussed later, in Sec.~\ref{sec:selfint}.
We initially focus on the operator $m_3^3$; as a check, in App.~\ref{app:spatflat} we perform the same calculation in spatially-flat gauge. 

\subsection{$m_3^3$-operator}
\label{subsec2.1}

Let us consider the action $S_0+S_{m_3}$.  One can restore the $\pi$ dependence in a generic gauge with the Stuekelberg procedure \cite{Cheung:2007st,Gleyzes:2013ooa} $t \rightarrow t+\pi(t,\vect{x})$. Focusing on the terms relevant for our calculations we have \cite{Cusin:2017mzw}
 \begin{align}
 g^{00} &\rightarrow g^{00}+2g^{0\mu}\partial_\mu\pi+g^{\mu\nu}\partial_\mu\pi\partial_\nu\pi \;,\label{eq:Stuekelberg1}\\ 
 \delta K &\rightarrow \delta K -h^{ij}\partial_i\partial_j\pi + \frac{2}{a^2} \partial_i \pi \partial_i \dot \pi +\ldots \; . \label{eq:Stuekelberg2} 
 \end{align}
In Newtonian gauge, the line element reads
 \begin{equation}
 \textrm ds^2 = -(1+2\Phi)\textrm dt^2 + a^2(t)(1-2\Psi)(e^\gamma)_{ij}\textrm dx^i \textrm dx^j \ ,
 \end{equation}
 with $\gamma$ transverse, $\partial_i \gamma_{ij} =0$, and traceless, $\gamma_{ii} =0$.
 
Varying the above action with respect to $\Phi$ and $\Psi$ and focussing on the sub-Hubble limit by keeping only the leading terms in spatial derivatives, one obtains
\be\label{PhiPsieq}
2 \MP^2 \nabla^2 \Psi + m_3^3   \nabla^2 \pi = 0 \;, \qquad  \MP^2 \nabla^2 (\Phi - \Psi)   =0\;,
\ee
where $\nabla^2 \equiv \partial_x^2  +  \partial_y^2 + \partial_z^2$. From now on we will always consider the Minkowski limit, i.e.~that time and spatial derivatives are much larger than Hubble.   
These equations can be solved in terms of $\pi$,
\be
\label{phisol}
\Phi = \Psi  = - \frac{m_3^3 }{2 \MP^2}  \pi \;.
\ee
Using these relations, one can find the kinetic term of the $\pi$ field. 

To write this, it is convenient to define the dimensionless quantity \cite{Bellini:2014fua,Gleyzes:2014rba}
\be\label{alphaB}
\alpha \equiv \frac{4 \MP^2 (c+ 2 m_2^4)+ 3 m_3^6}{2 \MP^4 H^2}   
 \;, 
  \ee
 which sets the normalization of the scalar fluctuations (and must be positive  to avoid ghost instabilities).
The canonically normalised scalar and tensor perturbations are then given by
 \begin{equation}\label{canonical}
 \pi_c \equiv  \sqrt{\alpha} \MP H \pi \;,\qquad \gamma_{ij}^{c} \equiv \frac{\MP}{\sqrt{2}}\gamma_{ij} \;.
 \end{equation}
In terms of these, the interaction term, after integrating by parts, reads
 \begin{equation} \label{InterationNew}
-\frac{1}{2}\sqrt{-g}~m_3^3\delta K\delta g^{00} \supset \frac{1}{\Lambda^2}\dot{\gamma}_{ij}^c\partial_i\pi_c\partial_j\pi_c \;,
 \end{equation} 
 where
 \begin{equation}
 \Lambda^2 \equiv \frac{4\MP^2(c+2m_2^4) + 3m_3^6}{\sqrt{2}m_3^3\MP}  = - \frac{\alpha}{\sqrt{2} \alpha_{\rm B}} \frac{H}{H_0} \Lambda_2^2\ \label{InterScale} \;,
 \end{equation}
 and in the right-hand side we have defined the dimensionless quantity\footnote{To write the action \eqref{total_action}, we used a conformal transformation (possibly dependent on $X= (\partial_\mu \phi)^2$) to set to constant the effective Planck mass and to zero higher-order operators of DHOST theories \cite{Langlois:2015cwa,Crisostomi:2016czh}. Using the notation of \cite{Langlois:2017mxy} and the transformation formulae contained therein, one can check that in a general frame the only relevant parameter is	
 \be
 \alpha_{\rm B} - \frac{\alpha_{\rm M}}{2} (1-\beta_1) + \beta_1 - \dot \beta_1 /H \;.
 \ee
}
\be\label{alphaB}
 \alpha_{\rm B} \equiv  -\frac{m_3^3}{2 \MP^2 H} \;.
\ee

We drop the symbol of canonical normalisation: $\gamma$ and $\pi$ will indicate for the rest of the paper the canonically normalised fields. The total Lagrangian is then
  \begin{equation}
  \label{InteractionLagrangian}
    S_0 +S_{m_3} = \int \text d^4x \left[\frac{1}{4}(\dot{\gamma}_{ij})^2 - \frac{1}{4}(\partial_k\gamma_{ij})^2 + \frac{1}{2}\dot{\pi}^2 - \frac{c_s^2}{2}(\partial_i\pi)^2 + \frac{1}{\Lambda^2}\dot{\gamma}_{ij}\partial_i\pi\partial_j\pi \right] \; ,
\end{equation}
 and 
 \begin{equation}
 c_s^2 = \frac{4\MP^2c-m_3^3(m_3^3-2\MP^2H)}{4\MP^2(c+2m_2^4) + 3m_3^6}   = \frac{2}{\alpha} \left( \frac{c}{\MP^2H^2} - \alpha_{\rm B}- \alpha_{\rm B}^2 \right) \;. \label{Cs}
 \end{equation}

The perturbative decay rate of the graviton can be computed following exactly the same procedure followed for $\tilde m_4^2$ in \cite{Creminelli:2018xsv}. It gives
\be
\Gamma_{\gamma\rightarrow\pi\pi} = \frac{p^5(1-c_s^2)^2}{480\pi c_s^7\Lambda^4} \ , \label{DecatRate}
\ee 
where $p$ is the momentum of the decaying graviton. 
One can check that this is negligible for frequencies relevant for GW observations, since by eq.~\eqref{InterScale} $\Lambda$ is of order $\Lambda_2 $.

The equation of motion of $\pi$ from the Lagrangian \eqref{InteractionLagrangian} reads
\begin{equation}
 \ddot{\pi} - c_s^2\nabla ^2\pi + \frac{2}{\Lambda^2}\dot{\gamma}_{ij} \partial_i\partial_j\pi = 0 \label{EoM1} \;.
 \end{equation} 
 Let us use the classical background solution of the GW travelling in the $\hat z$ direction with a linear polarization. Without loss of generality we take the $+$ polarization (the $\times$ one can be obtained by a $45^{\rm o}$ rotation of the axes)
\begin{equation}
\label{solutionh}
\gamma_{ij} = \MP h^+ \epsilon_{ij}^{+} \;, \qquad h^+ (t,z) \equiv h_0 ^+\sin(\omega(t-z))  \;,
\end{equation}
where $h^+$ is the dimensionless strain of the GW and the polarisation tensor is $\epsilon_{ij}^+ = {\rm diag}(1,-1,0)$. In this paper we will always be away from the source generating the GW, i.e.~in the weak field regime $h^+ \ll1 $.
Substituting the solution \eqref{solutionh} into eq.~\eqref{EoM1}, one gets (modulo an irrelevant phase)
\begin{equation}
\ddot{\pi} - c_s^2\nabla^2\pi + c_s^2 \beta  \cos[ \omega(t-z)] (\partial_x^2-\partial_y^2)\pi = 0 \; , \label{mastereq}
\end{equation}
where the parameter $\beta$ is defined as
\be
\label{betadef1}
\beta \equiv  
    \frac{2\omega \MP h_0^+}{c_s^2 | \Lambda^2 |} = \frac{2 \sqrt{2} | \alpha_{\rm B}| }{\alpha c_s^2  } \frac{\omega }{H} h_0^+  \;,  \qquad \text{for} \quad m_3^3 \neq 0 \;,  \ \tilde m_4^2 =0 \qquad (\alpha_{\rm B} \neq 0 \;,  \ \alpha_{\rm H} =0)  \;.
\ee
In the following, to evaluate the right-hand side of the above definition we will use $\omega = \Lambda_3$ and $H=H_0$, in which case $\omega/H \sim 10^{20}$.  Moreover, note that  $\alpha$ and $c_s^2$ in the second equality   appear in the combination $\alpha c_s^2$. Therefore, thanks to eqs.~\eqref{Cs}, $\beta$ defined above  depends only on $\alpha_{\rm B}$ (or $m_3$) and we can tune this parameter to make $\beta$ small.

\subsection{\texorpdfstring{$\tilde m_4^2$}{tilde m42}-operator}
\label{tildem42op}

We now consider the operator $\tilde m_4^2$. To avoid large $\pi$ nonlinearities, discussed in Sec.~\ref{sec:selfint}, we focus on the case $m_3 = 0$ and consider the action $S_0 + S_{\tilde m_4}$.
The operator $\tilde m_4^2$ has been studied in \cite{Creminelli:2018xsv} and details on the calculations can be found there.

In this case, the action for the canonically normalised fields reads
  \begin{equation}
  \label{ddotgamma}
    S_0 +S_{\tilde m_4} = \int \text d^4x \left[\frac{1}{4}(\dot{\gamma}_{ij})^2 - \frac{1}{4}(\partial_k\gamma_{ij})^2 + \frac{1}{2}\dot{\pi}^2 - \frac{c_s^2}{2}(\partial_i\pi)^2 + \frac{1}{\Lambda_\star^3}\ddot{\gamma}_{ij}\partial_i\pi\partial_j\pi \right] \; ,
\end{equation}
where the sound speed of scalar fluctuations is now 
\be
\label{soundsH}
c_s^2 = \frac{2}{\alpha} \left[(1+\alpha_{\rm H})^2  \frac{c}{\MP^2 H^2} + \alpha_{\rm H} + \alpha_{\rm H} (1+\alpha_{\rm H}) \frac{\dot H}{H^2} \right] \;,
\ee
and we have defined $\alpha_{\rm H} \equiv 2 \tilde m_4^2/\MP^2$ (we assumed $\alpha_{\rm H} \ll 1$ since this will be the regime of interest).
Thus, this operator does not affect the speed of propagation of GWs \cite{Creminelli:2017sry}, but, as shown in \cite{Creminelli:2018xsv}, it contains an interaction
$\gamma \pi \pi$ suppressed by the scale 
\begin{equation}
\Lambda_\star^3 \equiv \frac{\sqrt{2} \MP^3 (c + 2 m_2^4)}{ \tilde{m_4}^2(\MP^2 + 2 \tilde{m}_4^2)} \simeq \sqrt{2 }\frac{\alpha }{\alpha_\textrm{H}} \left( \frac{H}{H_0} \right)^2\Lambda_3^3 \;,
\end{equation}
where in the last equation we assumed a small $\alpha_{\rm H}$. This should be compared with the cubic coupling discussed above, eq.~\eqref{InteractionLagrangian}.

The evolution equation for $\pi$ then reads
\be
\ddot \pi - c_s^2 \nabla^2 \pi + \frac{2}{\Lambda_\star^3} \ddot \gamma_{ij} \partial_i \partial_j \pi =0 \;.
\ee
Substituting the solution \eqref{solutionh} into this equation gives
\begin{equation}
\ddot{\pi} - c_s^2\nabla^2\pi - \frac{2\omega^2 \MP h_0^+}{ \Lambda_\star^3}  \sin[ \omega(t-z)] (\partial_x^2-\partial_y^2)\pi = 0 \ . \label{EoM3}
\end{equation}
One sees that the evolution equation for $\pi$ for the $\tilde m_4^2$ operator is very similar to the case of the $m_3^3$ operator, with the replacement $\Lambda^2 \to \Lambda_\star^3 \omega^{-1}$. We can thus apply eq.~\eqref{mastereq}, with $\beta$ now defined as
\be    
\label{betadef2}
\beta \equiv  \frac{2 \omega^2 \MP h_0^+}{ c_s^2 | \Lambda_\star^3 |} = \frac{\sqrt{2} | \alpha_{\rm H}| }{\alpha c_s^2  } \left( \frac{\omega }{H}\right)^2 h_0^+   \;,  \qquad \text{for} \quad m_3^3 = 0 \;,  \ \tilde m_4^2 \neq 0 \qquad (\alpha_{\rm B} = 0 \;,  \ \alpha_{\rm H} \neq 0) \;.
\ee
Analogously to the $m_3$ case, because of eqs.~\eqref{soundsH}, $\beta$ defined  above
depends only on $\alpha_{\rm H}$ (or $\tilde m_4$) and also in this case we can tune this parameter to make $\beta$ small.

\section{Narrow resonance}
\label{sec:Resonance}

Equation \eqref{mastereq} describes a harmonic oscillator with periodic time-dependent frequency, which can lead to parametric resonance.
As explained in the introduction, in this article we are going to focus on the \emph{narrow-resonance} regime, which corresponds to $\beta \ll 1$.
In this case, the solution of the equation of motion of $\pi$ can be treated analytically and features an exponential growth of scalar fluctuations.

\subsection{Parametric resonance}
\label{subsec:ParametricResonance}

\begin{figure}[]
\centering 
\includegraphics[scale=0.13]{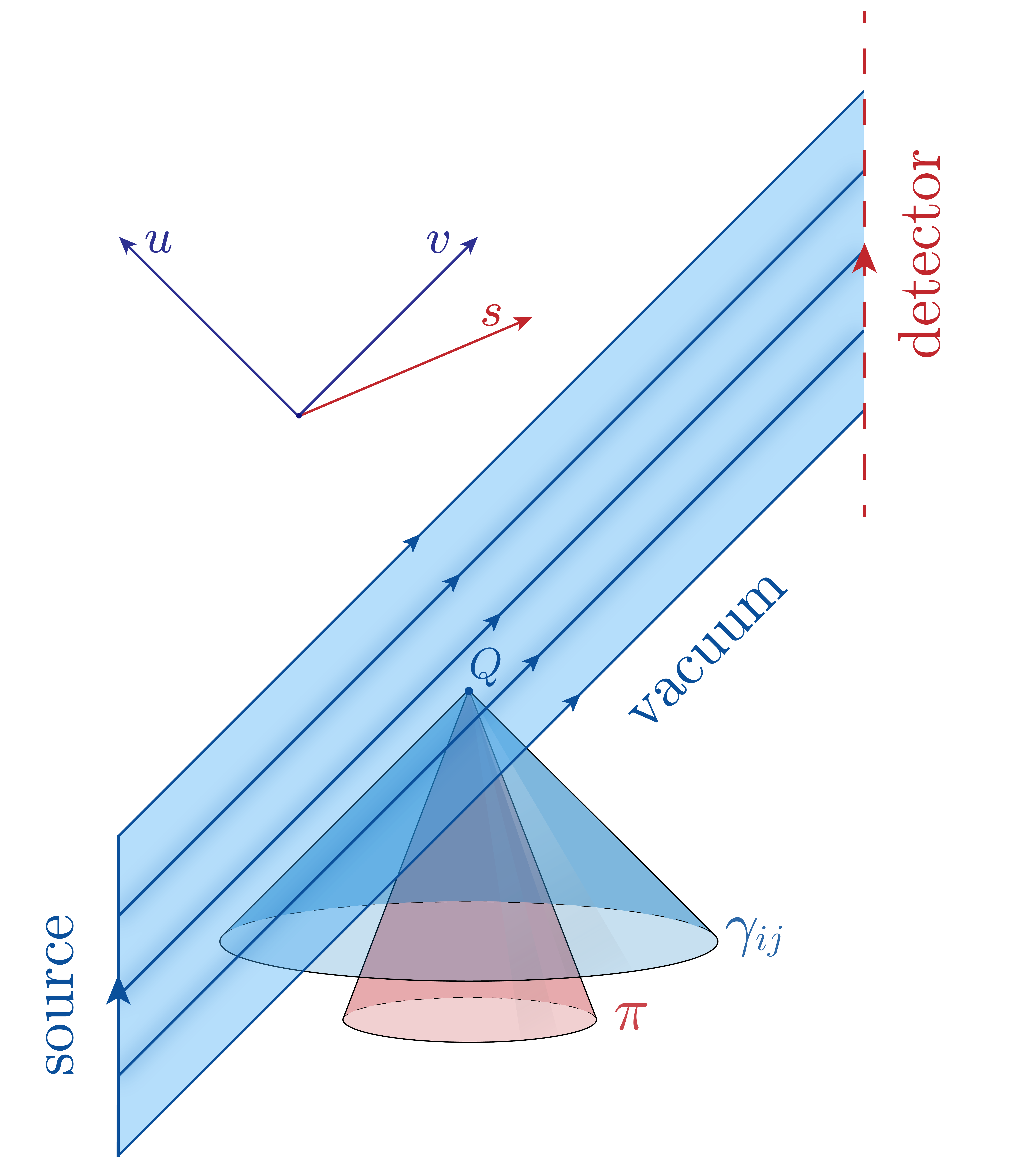}
\caption{\label{fig:diagram} Space-time diagram in $(t,z)$-coordinates indicating the path taken by the gravitational-wave wave packet (blue region). The $\pi$ light-cone is narrower than the gravitational wave one.}
\end{figure}

The GW is emitted at $t=0$ in the $z$ direction and is detected at some later time,  see Fig.~\ref{fig:diagram}. Using the light-cone coordinates,
\begin{equation}
u \equiv t-z\, , \qquad  v \equiv t +z \; ,
\end{equation}
its solution for $t>0$ can be written as
\begin{equation}
\gamma_{ij}(u) = \MP h_0^+  \sin\left( \omega u \right) \epsilon_{ij}^{+} \ ,
\end{equation}
where $h_0^+ $  can be taken as constant, since it varies slowly compared to the GW frequency.

Eq.~\eqref{mastereq} takes the form 
\be
\ddot \pi - c_s^2 \nabla^2 \pi +c_s^2  \betas  \cos\left( \omega u\right) (\partial_x^2  - \partial_y^2) \pi  = 0 \label{eq:1} \;.
\ee
For $c_s < 1$,  which we assume in the following, $u$ is a time-like variable for the $\pi$ metric: hypersurfaces of constant $u$ are space-like to the $\pi$-cone.
It is then convenient to define the variable $s$, 
\be
s \equiv -t + c_s^{-2} z \;,
\ee
which is orthogonal to $u$ and is thus space-like with the $\pi$ metric, see Fig.~\ref{fig:diagram}, and use the coordinates $\tilde{\vect x}= (x, y, s)$ to describe the spatial foliations. Since the $\pi$-lightcone is narrower than the one of GWs, the solution for the scalar will only be sensitive to a finite region of the GW background. This is the reason why one can approximate the GW as a plane wave with constant amplitude and disregard the process of emission of the GW from the astrophysical source.
\textcolor{black}{In particular, for this to be a good approximation we need to require that the past light-cone of $\pi$ overlaps only with a region of the GW with constant amplitude. From Fig.~\ref{fig:diagram} one sees that this is a very weak requirement. It is enough that $1- c_s$ is larger than the ratio between the duration of the GW signal (of order seconds) and the scale of variation of the amplitude (of order Mpc). Plugging the numbers one gets $1-c_s \gtrsim 10^{-14}$.} 

Since for $\pi$ there is translational invariance in $\tilde{\vect x}$, it is useful to  decompose $\pi$ in Fourier modes as
 \be
 \label{eq:Fourier}
 \pi (u, \tilde{\vect x}) = \int\frac{\text d^3 \tilde {\vect p} }{(2 \pi)^3}  e^{i \tilde{\vect p}\cdot \tilde{\vect x}} \pi_{\tilde{ \vect p}} (u)  \;,
 \ee
where $\tilde{\vect p}= (p_x, p_y, p_s)$ is conjugate to $\tilde{\vect x}$. In the absence of the interaction with the gravitational wave, one can relate $p_s$  to the momentum written in the original coordinates, 
\be\label{eq:ps}
p_s = \frac{c_s^2}{1-c_s^2}(p_z - c_s |\vect p|) \;.
\ee
In the following we are going to use this change of variable also when $\beta \neq 0$, although plane waves in the original coordinates $(t, x, y, z)$ are not solution of eq.~\eqref{eq:1}.

Then we can quantize $\pi$ straightforwardly.
More specifically, we  decompose $\pi_{\tilde{\vect p}} $ 
as
\be
\label{eq:pioperator}
\pi_{\tilde{ \vect p}} (u) = \frac{1}{c_s \sqrt{2 p_u}} \left[ f_{\tilde{ p}} (u) \hat a_{\tilde{ \vect p}}  + f^\star_{\tilde{ p}}(u) \hat a_{-\tilde{\vect{p}}}^\dagger \right] \;,
\ee
where 
\be
p_u \equiv  \frac{c_s}{1-c_s^2} (|\vect p| - c_s p_z ) \;,
\ee
and $\hat a_{\tilde{ \vect p}}$ and $\hat a_{-\tilde{\vect{p}}}^\dagger$ are the usual creation and annihilation operators satisfying the commutation relations, $[\hat a_{\tilde{\vect{p}}}, \hat a^\dagger_{\tilde{\vect{p}}'}] = (2 \pi)^3 \delta^{(3)}(\tilde{\vect p} - \tilde{\vect p}')$. 
The normalization is chosen for convenience. Indeed, for $\betas = 0 $ the evolution equation for $\pi$ satisfies a free wave equation and each Fourier mode can be described as an independent quantum harmonic oscillator. We assume that in this case $\pi$ is in the standard Minkowski vacuum, given by\footnote{It is straightforward to verify that eq.~\eqref{eq:vacuum} is equivalent to the standard Minkowski vacuum, i.e.
\be
\pi  (x) = \int \frac{ \text d^3 {\vect p}}{(2 \pi)^3} \frac{1}{\sqrt{2 c_s |{ \vect p}|}} \left( e^{-i  p \cdot  x} \hat a_{\vect p} + e^{i  p \cdot  x} \hat a^\dagger_{-\vect p}  \right) \;,
\ee
upon use of  $\text d^3 \tilde{\vect p} /\text d^3 \vect p  ={c_s p_u}/{|\vect{p}|}$ and, consequently, of $\hat a_{\tilde{ \vect p}} = \left[ {|{ \vect p}|}/({c_s p_u}) \right]^{1/2} \hat a_{{ \vect p}}$.}
\be
\label{eq:vacuum}
f_{\tilde{ p}} (u) =  e^{-i p_u u} \;, \qquad    \qquad (\betas=0)\;.
\ee

To study the parametric resonance,  will now show that  eq.~\eqref{eq:1} can be written as a Mathieu equation  \cite{mclachlan1951theory}.
First, in terms of the new coordinates, eq.~\eqref{eq:1} becomes
\begin{equation}\label{eq:eomp}
[(1-c_s^2)\partial_u^2 -c_s^{-2} (1-c_s^2) \partial_s^2 - c_s^2 (\partial_x^2 + \partial_y^2)] \pi + c_s^2 \beta \cos(\omega u) (\partial_x^2 - \partial_y^2)\pi = 0 \;.
\end{equation}
For convenience we can also define the dimensionless time variable $\tau$, 
\be
\tau \equiv \frac{\omega u}{2}  \;.
\ee
For each Fourier mode, $f$ satisfies 
\begin{equation}\label{eq:3}
\frac{\text d^2 f}{\text d \tau^2 }   + [A - 2 q \cos(2\tau)]f = 0 \;,
\end{equation}
with
\begin{align}
A &= 4 \frac{ c_s^2 \,\vect p^2 }{  \omega^2} \frac{(1-c_s \Omega)^2}{(1-c_s^2)^2} \label{eq:a} = \frac{4  p_u^2}{\omega^2}\;, \\
q &= 2 \betas \frac{ c_s^2  \, \vect p ^2  }{  \omega^2} \frac{(1-\Omega^2) \cos(2\varphi)}{1-c_s^2}\label{eq:q} \;.
\end{align}
To write $A$ and $q$ we have decomposed the vector $\vect p$ in polar coordinates, $\vect p = |\vect p|(\sin \theta \cos \varphi, \sin \theta \sin \varphi, \cos \theta)$, and we have defined $\Omega \equiv p_z/|\vect p| = \cos \theta$.

\begin{figure}[tbp]
\begin{subfigure}{0.5\textwidth}
\includegraphics[width=\linewidth]{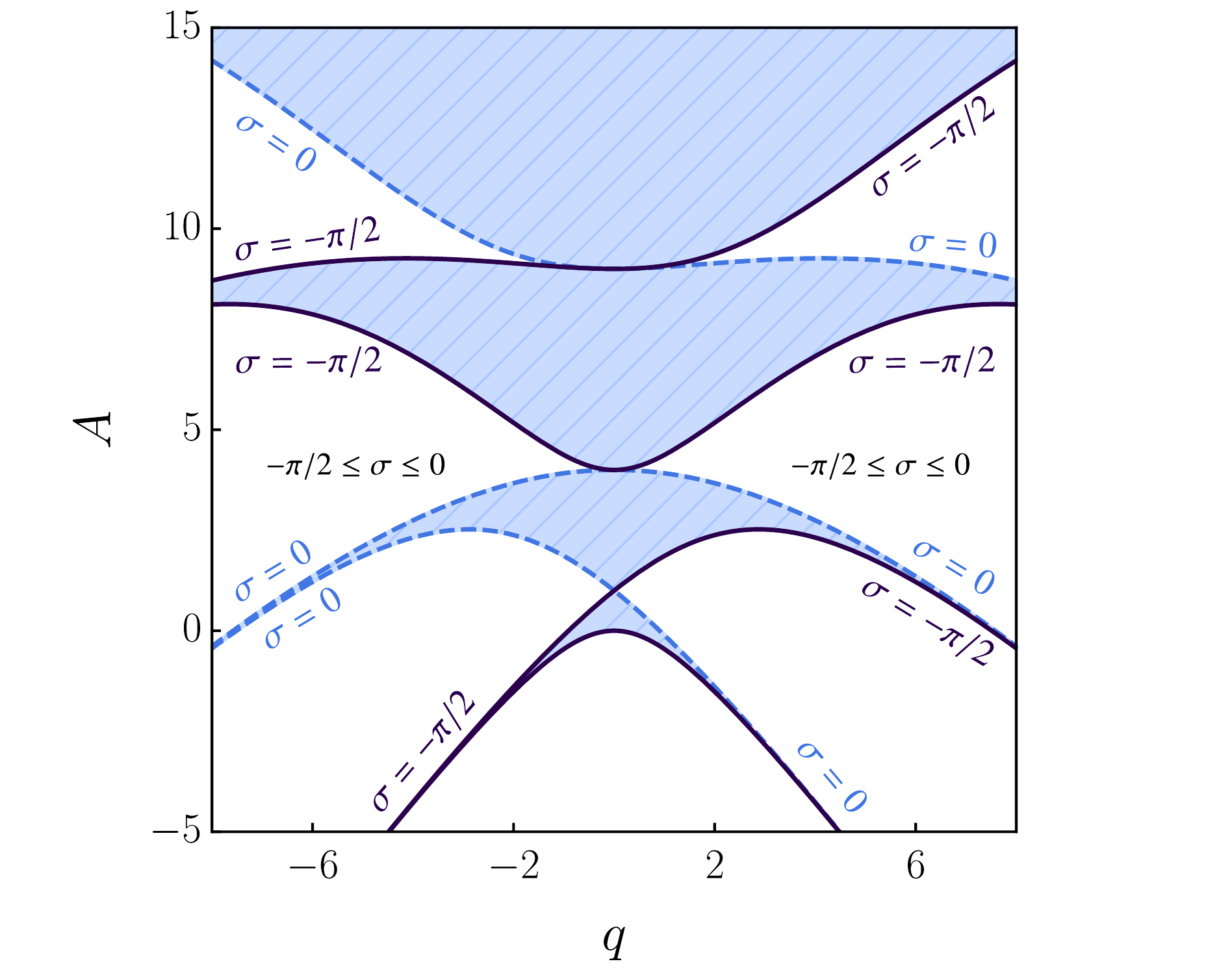}
\caption{\label{fig:mathsigmas}} 
\end{subfigure}
\hspace*{\fill} 
\begin{subfigure}{0.5\textwidth}
\includegraphics[width=\linewidth]{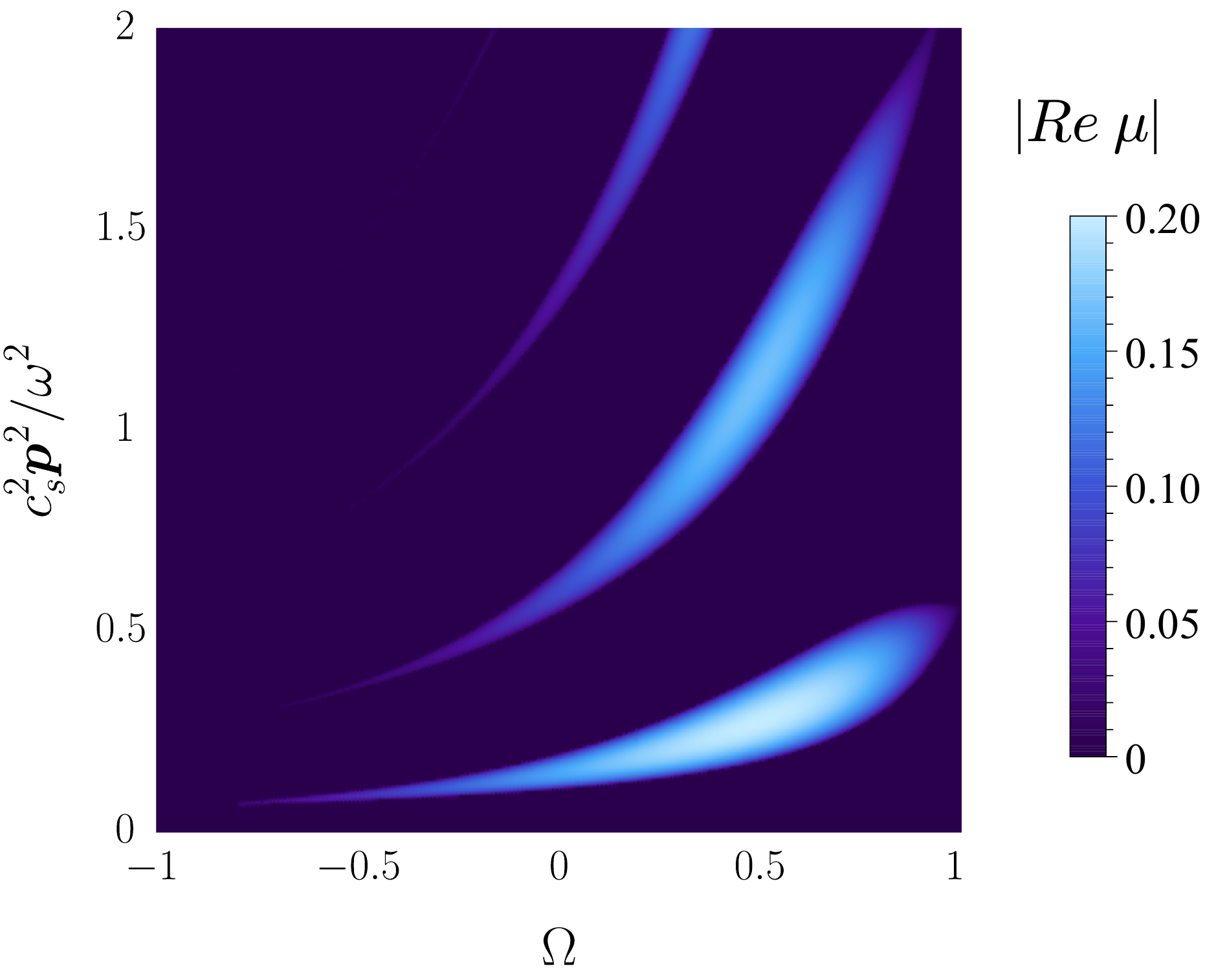}
\caption{\label{fig:math}} 
\end{subfigure}
\hspace*{\fill} 
\caption{~Left panel (Fig.~\ref{fig:mathsigmas}): Instability chart for the Mathieu equation \eqref{eq:3}. Colored regions are stable while empty regions are unstable. Each instability band is spanned by $\sigma \in \left(-\frac{\pi}{2}, 0 \right)$, see eq.~\eqref{eq:narrow_res_approx} and below for details. Right panel (Fig.~\ref{fig:math}): Instability bands as functions of $\Omega$ and $c_s^2 \vect p^2 / \omega^2$ assuming $\varphi=0$. Light blue regions are unstable: the color grading indicates the coefficient of instability $|Re\, \mu|$. The plot is obtained using $\beta = 0.8$ and $c_s = 1/2$. (We choose a large value of $\beta$ for this figure because the instability bands can be easily located, otherwise the bands would be too narrow to be seen.)  } 
\end{figure}

The general solution of the Mathieu equation is of the form $ e^{\pm \mu \tau} P(\tau)$, where $P(\tau + \pi) = P(\tau)$ \cite{mclachlan1951theory}.
If the characteristic exponent $\mu$ has a real part, the solution of the Mathieu equation is unstable for generic initial conditions. Since $\mu$ is a function of $A$ and $q$,  the instability region can be represented on the $(q,A)$-plane, see Fig.~\ref{fig:mathsigmas}.
The  unstable regions are also  shown in Fig.~\ref{fig:math} on the plane $(\Omega,{ c_s^2 \,\vect p^2 }/{  \omega^2})$, for specific values of the other parameters (in the example in the figure we take $\beta =0.8$ and $c_s = 1/2$).  Notice that the maximal exponential growth is reached when the ratio $q/A$ has its maximum at $\Omega  = c_s$ and $\varphi = 0$.

\subsection{Energy density of $\pi$}
\label{subsec:pienergy}

In this subsection we  compute the energy density of $\pi$ produced by the coupling with the GW.  
This is given by (see  eq.~\eqref{T_00} and explanation below)
\begin{align}
\rho_{\pi} &= \frac{1}{2} \langle 0| \left[\dot{\pi}^2 + c_s^2 (\partial_i \pi)^2  \right]  |0 \rangle \;.
\end{align}
Decomposing in Fourier modes and in  the $\hat a$ and $\hat a^\dagger$ coefficients using respectively eqs.~\eqref{eq:Fourier} and \eqref{eq:pioperator}, the energy density can  be rewritten as
\begin{align}
\label{eq:rhopi}
\nonumber \rho_{\pi} &= \int \frac{\text d^3 \tilde{\vect p}}{(2 \pi)^3 }\frac{1}{4 c_s^2  p_u}\left[ |f_{\tilde{p}}' - i p_s f_{\tilde{p}}|^2  + c_s^2 |f_{\tilde{p}}' - i c_s^{-2}p_s f_{\tilde{p}}|^2  + 2 i p_s  (f'_{\tilde{p}} f^\star _{\tilde{p}} - {f'}^{\star}_{\tilde{p}} f _{\tilde{p}})  + c_s^2 (p_x^2 + p_y^2)|f_{\tilde{p}}|^2 \right] \\
& = \int \frac{\text d^3 \tilde{\vect p}}{(2 \pi)^3 }\frac{1}{4 c_s^2  p_u} \left\{(1+c_s^2) | \partial_u f_{\tilde{p}}|^2  + |f_{\tilde{p}}|^2\left[ p_s^2 (1+c_s^{-2})  + c_s^2 (p_x^2 + p_y^2)  \right]   + 4  p_s p_u  \right\} \;,
\end{align}
where we have simplified the expression on the right-hand side using that the Wronskian is time-independent, $\mathscr W [f_{\tilde{p}}(u), f^\star_{\tilde{p}}(u)] = - 2 i p_u$.\footnote{In general, we can write $f$ as a linear combination of Mathieu-sine and cosine functions, respectively $\mathcal S$ and $\mathcal C$ \cite{mclachlan1951theory}. We fix the boundary conditions of the solution such that $\pi$ is in the vacuum, i.e.~the function $f$ satisfies eq.~\eqref{eq:vacuum}, at $u=0$. This gives
\begin{equation}\label{eq:4}
f(\tau) = \frac{-i  2 p_u }{ \omega }\, \mathcal S\left(A, q; \tau \right) +  \, \mathcal C\left(A, q; \tau \right) \;.
\end{equation}
Using this expression and the properties of these functions, we can check that the Wronskian $\mathscr W [f_{\tilde{p}}(u), f^\star_{\tilde{p}}(u)]$ is constant and with the above normalization is given by $-2 i p_u$. As a consequence, 
the commutation relation in the ``interacting'' region are satisfied at all times.
}
One can verify that in the limit $\beta = 0$ the above expression reduces to the   energy density of the vacuum, i.e.~$\rho_{\pi}^0 =  \int \frac{\text d^3 p }{(2 \pi)^3} \frac{\omega_p}{2} $.

We can simplify the right-hand side further by making some approximations.
Since we are not interested in following the oscillatory behaviour of $\rho_\pi$, we can perform an average in $\tau$ over many periods of oscillation. 
Since the amplitude of the periodic part of $f_{\tilde{\vect p}}$ is bounded to be less than unity, it is reasonable to take 
$\braket{|f_{\tilde{ p}}|^2}_T \simeq \braket{|\partial_\tau f_{\tilde{ p}}|^2}_T  \simeq  e^{2 \mu \tau} /2 $ in eq.~\eqref{eq:rhopi}. This educated guess will be confirmed in Sec.~\ref{subsec:waveformmod}. Changing integration variables, from $\tilde{\vect{p}}$ to  $\xi \equiv c_s^2 \vect{p}^2 / \omega^2$ and the angular variables $\Omega$ and $\varphi$, we find
\begin{align}
{\rho_{\pi}} \simeq \frac{\omega^4}{(2\pi)^3 16 c_s^3}\int \text d\Omega \, \text d \varphi \, \text d  \xi  \, \left[ \frac{1+c_s^2}{4}  + \xi \frac{(c_s- \Omega)^2 (1+c_s^{2}) }{(1-c_s^2)^2} +  \xi (1-\Omega^2) \right]
 e^{2 \mu (\xi, \Omega,\varphi) \tau}  \;.
 \label{eq:integral}
\end{align}

We now want to solve the integral on the right-hand side.
Since it is dominated by the unstable modes, we restrict the domain of integration to the bands of instability. Actually, we are going to focus on the first instability band for two reasons. 
\textcolor{black}{The instability rate of the higher bands goes as $\mu_m \sim q^m / (m!)^2 \sim \beta^m$ \cite{mclachlan1951theory,olver2010nist}, which implies that for small $\beta$ the first band is the most unstable.
Second,} the $\pi$ produced in this band will modify the GW signal with the original angular frequency $\omega$. (With some contributions at higher frequencies that will be discussed in Section \ref{sec:mtilde42}.)  We are going to work in the regime $\beta \ll 1$ corresponding to $q \ll 1$, the regime of narrow resonance.

In this situation we can restrict the integral to the first unstable band, which is defined by \cite{mclachlan1951theory}
\begin{equation}\label{eq:boundary}
A_- \le A \le A_+ \;, \qquad  A_\pm = 1 \pm |q| \;.
\end{equation}
Within this region, the value of the exponent $\mu$ is 
\begin{equation}
\label{eq:muNR}
 \mu \simeq \frac{1}{2}\sqrt{(A_+ - A)(A- A_-)}\,.
 \end{equation} 
Using the definitions \eqref{eq:a} and \eqref{eq:q} in eq.~\eqref{eq:boundary}, the boundary region above can be rewritten in terms of $\xi$, 
\begin{equation}
\xi_-\le \xi \le \xi_+ \;, \qquad \xi_\pm= \frac{(1-c_s^2)^2}{4(1-c_s \Omega)^2}\left[ 1 \pm   \beta |\cos ( 2\varphi) | \frac{(1-\Omega^2) (1-c_s^2)}{2  (1-c_s \Omega)^2}\right] \,,
\end{equation} 
which fixes the domain of integration in eq.~\eqref{eq:integral}.

The integral in  \eqref{eq:integral} can be then solved  with the saddle-point approximation. In general, an integral of the form
\begin{equation}
I(\tau) = \int \text d ^3 \vect X g(\vect X) e^{ h(\vect X) \tau }  \;,
\end{equation}
can be approximated for large $\tau$ by
\begin{equation}\label{eq:saddle_approx}
I(\tau)\approx  \frac{1}{\sqrt{\text{det} \,{\rm H}_{ij}}} \left(\frac{2 \pi}{\tau} \right)^{3 / 2} g(\vect X_0)e^{h(\vect X_0)\tau } \;, 
\end{equation}
where $\vect X_0$ is the maximum of the exponent, i.e.~the solution of $\partial_i h(\vect X) = 0$ with the Hessian  ${\rm H}_{ij} \equiv -\partial_i \partial_j h (\vect X_0)$ positive definite. (If there are more than one maximum one should sum over them.)

In our case we have to maximize  $ \mu$ of eq.~\eqref{eq:muNR}, by requiring that $\partial \mu / \partial \xi$, $\partial \mu/ \partial \Omega $ and $\partial \mu/ \partial \varphi$ vanish.
This happens for  
\be
\label{eq:solSP}
\xi =  \frac1{4 -\betas^2}\, , \qquad \Omega = c_s \;, \qquad \varphi = n \pi/4 \;, \qquad (n=0,1,2,\ldots) \;,
\ee
but we discard the solutions with $n$ odd, because they make $q$ (and the integration region) vanish, see eq.~\eqref{eq:q}. \textcolor{black}{There are four relevant saddle points $(\varphi = 0, \pi/2, \pi, 3\pi/2)$ giving the same value to the integral}; without loss of generality we choose $\varphi=0$. 
The expression of the Hessian matrix at the saddle point is
\begin{equation}\label{eq:matrixH}
\text H_{ij} = 4 \left(
\begin{array}{ccc}
 \frac{\left(4-\beta ^2\right)^{3/2} }{\beta } & -\frac{2 c_s \sqrt{4-\beta ^2} }{\beta (1-c_s^2)
 } & 0 \\
 -\frac{2 c_s \sqrt{4-\beta ^2} }{\beta (1-c_s^2)} & \frac{4 c_s^2 \left(4-\beta
   ^2\right)+2 \beta ^2 }{\left(1- c_s^2\right)^2 \beta  \left(4-\beta ^2\right)^{3/2}} & 0 \\
 0 & 0 & \frac{4 \beta }{\left(4-\beta ^2\right)^{3/2}} \\
\end{array}
\right)\;.
\end{equation}

Working at leading order in $\beta $ and using the saddle-point approximation with these values in eq.~\eqref{eq:integral},
the energy density of $\pi$ as a function of $u=t-z$ reads
\begin{equation}
\label{eq:pienergy}
{\rho_{\pi}} (u)\approx \frac{\omega^{5/2}(1-c_s^2) }{c_s  ( 8 \pi u)^{3/2} \sqrt{\beta} } \exp\left( \frac{\betas}{4} \omega u \right)\;.
\end{equation}
We can  compare $\rho_\pi$ with the energy density of the gravitational wave, which is roughly constant,
\begin{equation}
{\rho_\gamma} \simeq (\MP  \omega  h_0^+)^2\,.
\end{equation}
For instance, for $\beta = 0.1$ and $c_s = 1/2$, we get ${\rho_{\pi}} \simeq \rho_\gamma$ after $\tau/\pi \simeq 750$ cycles. 

The exponential growth studied in this section can also be seen as a consequence of Bose enhancement, see Appendix \ref{Bose}.

\subsection{Modification of the gravitational waveform}
\label{subsec:waveformmod}
The parametric production of $\pi$  suggests that its back-reaction  will modify the background gravitational wave. In this section we estimate this effect remaining in the narrow-resonance limit, i.e.~$|q|\ll1$ ($\beta \ll1$). Here we focus again on the case of the operator $m_3^3$.

To compute the  back-reaction on $\gamma_{ij}$, we start from the action \eqref{InteractionLagrangian}.
The equation of motion for $\gamma_{ij}$ is 
\begin{equation}
\ddot \gamma_{ij} - \nabla^2 \gamma_{ij} +\frac{2}{\Lambda^2}  \Lambda_{ij,kl}  \partial_t \left(\partial_k \pi \partial_l \pi\right) = 0 \; ,
\end{equation}
where, given a direction of propagation of the wave $\vect n$,   $\Lambda_{ij,kl}(\vect n)$ is the projector into the traceless-transverse gauge, defined by 
 \be
 \Lambda_{ij,kl} (\vect n)  \equiv  (\delta_{ik}-n_in_k)(\delta_{jl}-n_jn_l)-\frac{1}{2}(\delta_{ij}-n_in_j)(\delta_{kl}-n_kn_l) \;.
 \ee
We focus again on a wave traveling in the $\hat{ \vect z}$ direction. Using light-cone variables the equation above becomes
\begin{equation}
\label{eq:evolgamma}
\partial_u \partial_v \gamma_{ij} +\frac{1}{4 \Lambda^2} \partial_u   J_{ij} (u)  = 0 \; , \qquad J_{ij} (u)  \equiv  \Lambda_{ij,kl} \partial_k \pi \partial_l \pi  = \Lambda_{ij,kl} \int \frac{\text d ^3 \tilde {\vect p} }{(2\pi)^3} \frac{ 2  p_k p_l}{  c_s^2 p_u }|f_{\tilde{ p}}(u)|^2 \; .
\ee
We can then split the solution into a homogeneous and a forced one, 
\be
\label{eq:gammasplit}
\gamma_{ij} \equiv \bar \gamma_{ij} + \Delta \gamma_{ij}\;. 
\ee
The former reads
\be
\bar \gamma_{ij}(u,v) = \bar \gamma_{ij}(u, 0) +\bar  \gamma_{ij}(0, v) - \bar  \gamma_{ij}(0, 0) \;,
\ee
while for the latter, which represents the back-reaction due to $\pi$, we find
\be
\Delta \gamma_{ij} (u,v)= -\frac{1}{4 \Lambda^2}   \left[   J_{ij} (u)  -J_{ij} (0)  \right]   v \;. \label{eq:sol_backreact}
\ee

Because it is transverse and traceless, the  source $J_{ij}$ can be projected into a plus and cross polarization. We will focus on the plus polarization. In this case, we can proceed as in the previous subsection and rewrite the integral in terms of $\Omega$, $\varphi$ and $\xi$, 
\be
\label{eq:intJ}
J_{ij} (u) =\frac{\omega^4 }{4 (2\pi)^3 c_s^5  } \int \text d \Omega \, \text d \varphi\,   \text d \xi  \;  \xi  \, (1-\Omega^2)  \cos (2 \varphi)  |f_{\tilde{ p}}(u)|^2 \epsilon_{ij}^{+}  \;.
\ee
Then, we can evaluate this integral in the narrow-resonance regime with the saddle-point approximation. However,  we will now use an approximation for the solution $f_{\tilde{ p}}(u)$ that takes into account  its oscillatory behaviour. 
In particular, we will split the integral in two regions: $q>0$ and $q<0$. Since the sign of $q$ is controlled  by the angle $\varphi$ through $\cos 2 \varphi$, this  corresponds to splitting the integration over $\varphi$.

For $q>0$, the solution of the Mathieu equation \textcolor{black}{in the first instability band} can be  approximated  by (see pag. 72 of \cite{mclachlan1951theory})
\begin{equation}\label{eq:narrow_res_approx}
f(\tau) \simeq c_+ \, e^{\mu \tau} \sin(\tau - \sigma) + c_-\, e^{-\mu \tau} \sin(\tau +\sigma)\;, \qquad (q>0) \;, 
\end{equation}
where $\mu>0$ and $\sigma \in ( -\frac{\pi}{2}, 0 )$ is a  parameter, which depends on $A$ and $q$. It is real inside the instability bands, as shown in the instability chart for the Mathieu equation in Fig.~\ref{fig:mathsigmas}. More specifically, in the first instability band  one has 
\begin{align}
A &= 1 - q \cos(2\sigma) + \mathcal O(q^2) \label{eq:a_sigma} \;, \\
\mu &= -\frac{1}{2} q \sin(2 \sigma) + \mathcal O(q^2)\label{eq:mu_sigma} \;.
\end{align}
The coefficients $c_+$ and $c_-$ can be fixed by demanding that at $\tau = 0$ we recover the vacuum solution. 

The case $q<0$ can be obtained by noting that when $q$ changes sign, $\mu$  does it as well while  $A$ remains the same. The only way to implement this is to consider the simultaneous change $\sigma \rightarrow \sigma'  = -\sigma - \frac{\pi}{2}$. By performing these two transformations for $q$ and $\sigma$ on $f(\tau)$ one obtains
\begin{align}\label{eq:narrow_res_approx2}
f(\tau) \simeq c_-' e^{-\mu \tau} \cos(\tau + \sigma) - c_+' e^{\mu \tau} \cos(\tau - \sigma) \;, \qquad (q<0) \;.
\end{align}

We can now integrate the right-hand side of \eqref{eq:intJ} starting from  the interval $\varphi \in \left( - \frac{\pi}{4}, \frac{\pi}{4}\right )$. Replacing the growing mode solution of eq.~\eqref{eq:narrow_res_approx} in the integrand, we obtain
\begin{align}
 \frac{\omega^4 }{4 (2 \pi)^3c_s^5 }\int_{-\pi/4}^{\pi /4} \text d \varphi\,\int_{-1}^{1} \text d \Omega \, \int \text d \xi \, \xi  \, (1-\Omega^2)\cos ( 2 \varphi)  |c_1|^2 \sin^2 (\tau - \sigma)e^{2 \mu \tau} \label{eq:J1} \;.
\end{align}
The $\tau$ dependence in $\sin(\tau -\sigma)$ seems to change the saddle point computed by maximizing $\mu$. However, by rewriting the sine as exponential functions one can check that its effect is simply to add an additional constant to the exponent so that the saddle point remains the same as the one we computed in Sec.~\ref{subsec:pienergy}, see eq.~\eqref{eq:solSP}. 
At the saddle point, eqs.~\eqref{eq:a_sigma} and \eqref{eq:mu_sigma} give $\tan (2 \sigma) \simeq 2/\beta$ while $c_+$ and $c_+'$ can be computed by requiring vacuum initial conditions, eq.~\eqref{eq:vacuum}, at $\tau=0$. Working for small $\beta$, this gives
\be
\label{eq:SPscc}
\sigma \simeq -\frac{\pi + \beta}{4} \;, \qquad c_+ \simeq c_+' \simeq  \frac{1 - i}{\sqrt 2}\;.
\ee

Then, applying eq.~\eqref{eq:saddle_approx} to the integral in \eqref{eq:J1}, the latter can be solved, giving
\be
\frac{\omega^4 (1-c_s^2)^2  }{2 c_s^5 (16 \pi \tau)^{3/2} \sqrt{  \beta}} | c_{+}|^2 \sin^2 (\tau - \sigma)e^{ \beta \tau/2}  \;, 
\ee
where for $c_+$ and $\sigma$ we must use the saddle-point values, eq.~\eqref{eq:SPscc}.
We can now repeat this exercise for each part of the integral, using  the growing mode of either eqs.~\eqref{eq:narrow_res_approx} or \eqref{eq:narrow_res_approx2} depending on the sign of $q$. Summing them together, we obtain
 \begin{equation}\label{eq:fullJ}
  J_{ij} \simeq  \frac{\omega^4 (1-c_s^2)^2}{c_s^5 (16 \pi \tau)^{3/2}  \sqrt{ \beta}} \,  \left[ |c_+|^2 \sin^2 (\tau - \sigma )- |c_+'|^2 \cos^2 (\tau - \sigma )  \right]   e^{\beta \tau/2} \epsilon_{ij}^+ \; .
 \end{equation}

Replacing in the solution for $\Delta \gamma_{ij}$, eq.~\eqref{eq:sol_backreact}, the expression of 
$J_{ij}(u)$ of eq.~\eqref{eq:fullJ}, with $\sigma$, $c_+$ and $c_+'$ fixed by the saddle-point approximation from eq.~\eqref{eq:SPscc}, the  back-reaction on the GWs due to $\pi$ reads 
\begin{equation}\label{eq:GammaBR}
\Delta \gamma_{ij} (u, v) \simeq -\frac{v}{4 \Lambda^2}  \frac{  (1-c_s^2)^2}{  c_s^5  \sqrt{ \beta}} \,   \frac{\omega^{5/2}}{(8  u \pi )^{3/2}}    \sin \left( \omega u + \frac{\beta}{2} \right)  \exp \left( \frac{\beta}{4} \omega u \right) \epsilon_{ij}^+  \;,
\end{equation}
where we have dropped $J_{ij}(0)$ which is negligible at late time. 
Thus, the back-reaction  grows exponentially in $u$, as expected from the growth of the energy of $\pi$, eq.~\eqref{eq:pienergy}, and  linearly in $v$. Note that the right-hand side of the above equation diverges for $\beta \rightarrow 0$. This is because it has been obtained using the saddle-point approximation, which assumes that $\beta \omega u$ is large. In Appendix \ref{Bose} we check that this result reduces to the perturbative calculation when the occupation number is small enough.

Notice that there is no production of cross polarization.  Indeed, the integrand of the source term in eq.~\eqref{eq:evolgamma} now contains $2 p_x p_y$ instead of $p_x^2 - p_y^2$. Since $p_x p_y \propto \sin (2 \varphi) = 0 $
and the saddle points are such that $\sin (2 \varphi) = 0$, the cross-polarized waves are not generated by the back-reaction of dark energy fluctuations produced by plus-polarized waves and eq.~\eqref{eq:GammaBR} represents the full result.

\subsection{Generic polarization\label{sec:genpol}}
So we far we have been discussing linearly polarized waves. Since the resonant effect is non-linear, one cannot simply superimpose the result for linear polarization in order to get a general polarization.
In this subsection we are going to consider a more generic polarization state, that we parametrise as follows
\be
\gamma_{ij} = \MP h_0 \left[ \cos\alpha\sin(\omega u) \epsilon_{ij}^+ + \sin\alpha \cos(\omega u)\epsilon_{ij}^\times\right] \;, 
\ee
where $0 \leq \alpha< 2\pi$ is an angle characterizing the GW polarization.
Note that the state of polarization for generic $\alpha$ is \emph{elliptical}, like the one coming from binary systems \cite{Maggiore:1900zz}. 

Following the same procedure as in Sec.~\ref{subsec:ParametricResonance}, the Mathieu eq.~(\ref{eq:3}) becomes
\be
\frac{\text d^2 f}{\text d \tau^2 }   + [A - 2 q \cos(2\tau + \hat{\theta})]f = 0 \;, \qquad \tan\hat{\theta} \equiv \tan\alpha \tan(2\varphi) \; ,
\ee
with
\be 
q = \sqrt{2} \beta  \frac{ c_s^2  \, \vect p ^2  }{  \omega^2} \frac{(1-\Omega^2)}{1-c_s^2} \sqrt{ 1+ \cos (2\alpha)\cos(4\varphi) } \label{eq:q_generic} \;,
\ee
while $A$ remains the same as before. One needs to shift $\tau \rightarrow \tau + \hat{\theta}/2$ in order to use the same form for the Mathieu solution. Given this change in $q$, one can easily obtain the modified saddle points which are given by 
\begin{align}
\xi =  \frac1{4 -\frac{\betas^2}{2}[1 + (-1)^n\cos(2\alpha)]}\, , \qquad \Omega = c_s \;, \qquad \varphi = n \pi/4 \;, \qquad (n=0,1,2,\ldots) \;.
\end{align}
Thus, the saddle points for $\Omega$ and $\varphi$ are unaffected by the angle $\alpha$. From the saddle point found above one can see that choosing $n$ to be even selects the $+$ polarization, whereas $n$ odd corresponds to $\times$ polarization. 
The exponent $\mu$ of eq. (\ref{eq:muNR}) can be evaluated on these saddle points and, at leading order in $\beta$, is given by
\begin{align}
\mu \simeq \frac{\beta}{4\sqrt{2}}\sqrt{1 + (-1)^n\cos(2\alpha)} \;. 
\end{align}
Here one can define $\mu^+$ and $\mu^\times$ corresponding to $+$ and $\times$ polarizations respectively as 
\begin{align}
\mu^+ \equiv \frac{\beta}{4\sqrt{2}}\sqrt{1 + \cos(2\alpha)}\, , \qquad \mu^\times \equiv \frac{\beta}{4\sqrt{2}}\sqrt{1 - \cos(2\alpha)} \;.
\end{align}

Several comments are in order at this stage. First, for $0 < \alpha < \pi/4$ the $+$ contribution in the initial wave is larger. In this case one has $\mu^+ > \mu^\times$ which means that the $+$ polarization dominates also in the back-reaction for $\Delta \gamma_{ij}$. For $\pi/4 < \alpha < \pi/2$ the $\times$ mode dominates. For $\alpha = \pi/4$ both polarization states grow with the same rate, meaning that a circularly polarized wave remains circular. Moreover, by setting $\alpha = 0$ ($\alpha = \pi/2$) we recover the results of the previous sections for the case of $+$ ($\times $) polarization.

\subsection{Conservation of energy\label{sec:conservation}}
To check our results and to get a better understanding of the system, it is useful to verify that energy is conserved in the production of the $\pi$ field and the corresponding modification of the GW, $\Delta\gamma$. From the Lagrangian \eqref{InteractionLagrangian}, one can derive the Noether stress-energy tensor, which is conserved on-shell as a consequence of translational invariance
\be
\label{eq:emc}
\partial_\mu T^\mu_{\ \nu} = 0 \;.
\ee
(Notice that this $T^\mu_{\ \nu}$ will be different from the pseudo stress-energy tensor of GR.) Let us consider the region represented in Fig.~\ref{fig:diagram_boundary}. Given the symmetries of the system, it is useful to take the left and right boundaries as null, instead of time-like, surfaces.
\begin{figure}[h!]
\centering 
\includegraphics[scale=0.4]{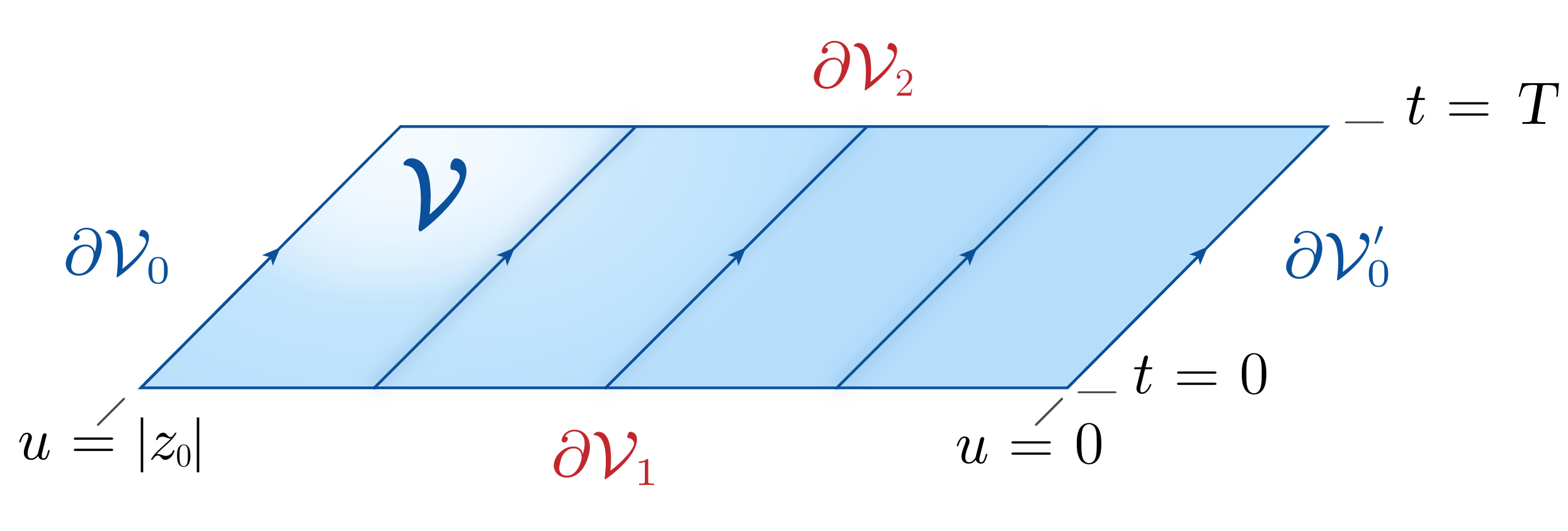}
\caption{\label{fig:diagram_boundary} Region $\mathcal V$ over which we are checking the conservation of the stress-energy tensor. The boundaries $\partial \mathcal V_0$ and $\partial \mathcal V'_0$ are null hypersurfaces at $u = const$.}
\end{figure}

At first sight, it is somewhat puzzling that while both the original GW and the induced $\pi$ are only displaced as time proceeds (see Figs.~\ref{fig:diagram} and \ref{fig:diagram_boundary}), so that their contribution to the total energy does not depend on time, $\Delta\gamma$ grows on later time slices since it is proportional to $v$. What we are going to verify is that the variation of the energy between $\partial\mathcal V_1$ and $\partial\mathcal V_2$ due to the change in $\Delta \gamma$ is compensated by a flux of energy across the null boundary $\partial\mathcal V_0$.\footnote{Notice that, while the original GW is described by a wave packet localised in a certain interval of $u$, $\pi$ waves are present at arbitrary large $u$ and thus will always contribute to the flux.} (There is no flux across the right null boundary $\partial\mathcal V_0'$ since $\pi$ is in the vacuum for $u=0$.)

The Noether stress-energy tensor of the action \eqref{InteractionLagrangian} is given by
\begin{align}
    T^0_{\ 0} &= (T^0_{\ 0})_\gamma + (T^0_{\ 0})_\pi \;, \qquad (T^0_{\ 0})_\gamma \equiv -\frac{1}{4}\left[ \dot{\gamma}^2_{ij} + (\partial_k\gamma_{ij})^2 \right]  \;, \qquad (T^0_{\ 0})_\pi \equiv- \frac{1}{2} \left[ \dot{\pi}^2 + c_s^2(\partial_i\pi)^2 \right] \; \label{T_00}, \\
   T^0_{\ i} &= -\frac{1}{2}\dot{\gamma}_{kl}\partial_i\gamma_{kl} - \dot{\pi}\partial_i\pi - \frac{1}{\Lambda^2}\partial_i\gamma_{kl}\partial_k\pi\partial_l\pi \; \label{T_0i} , \qquad
   T^i_{\ 0} = \frac{1}{2}\dot{\gamma}_{kl}\partial_i\gamma_{kl} + c_s^2\dot{\pi}\partial_i\pi - \frac{2}{\Lambda^2}\dot{\gamma}_{ij}\dot{\pi}\partial_j\pi \;   , \\
   T^i_{\ j} &= \frac{1}{2}\partial_i\gamma_{kl}\partial_j\gamma_{kl} + c_s^2\partial_i\pi\partial_j\pi  - \frac{2}{\Lambda^2}\dot\gamma_{ik}\partial_j\pi\partial_k\pi  \nonumber \\
   & \quad + \frac{1}{2}\delta^i_j\left[\frac{1}{2}\dot{\gamma}_{kl}^2 - \frac{1}{2}(\partial_m\gamma_{kl})^2 + \dot{\pi}^2 -c_s^2(\partial_l\pi)^2 + \frac{2}{\Lambda^2}\dot{\gamma}_{kl}\partial_k\pi\partial_l\pi \right] \; \label{T_ij}. 
\end{align}
Notice that the total energy of the system is simply the sum of the kinetic energy of $\gamma$ and $\pi$ without a contribution due to interactions. (This is a consequence of the interaction term being linear in $\dot\gamma$.) This means that the production of $\pi$ must be compensated by a decrease of the $\gamma$ kinetic energy.
Using the splitting in eq.~\eqref{eq:gammasplit} and defining $\bar \rho_\gamma \equiv \frac{1}{4}(\dot{\bar \gamma}_{ij})^2 + \frac{1}{4}(\partial_k \bar \gamma_{ij})^2$, $\rho_\pi \equiv \frac{1}{2}\dot{\pi}^2 + \frac{c_s^2}{2}(\partial_i\pi)^2$, the components \eqref{T_00} \eqref{T_ij} can be written up to second order in perturbation. For instance
\begin{align}
    T^0_{\ 0} &= -(\bar \rho_\gamma + \rho_\pi + \frac{1}{2}\dot{\bar\gamma}_{ij} {\Delta\dot \gamma}_{ij} + \frac{1}{2}\partial_k\bar\gamma_{ij}\partial_k\Delta\gamma_{ij}) + \mathcal{O}(\Delta\gamma^2) \; \label{perturb:T_00}\\
    T^i_{\ 0} &= \frac{1}{2}\dot{\bar\gamma}_{kl}\partial_i\bar\gamma_{kl} + \frac{1}{2}\Delta\dot{\gamma}_{kl}\partial_i\bar\gamma_{kl} + \frac{1}{2}\dot{\bar\gamma}_{kl}\partial_i\Delta\gamma_{kl} + c_s^2\dot{\pi}\partial_i\pi - \frac{2}{\Lambda^2}\dot{\bar\gamma}_{ij}\dot{\pi}\partial_j\pi + \mathcal{O}(\Delta\gamma^2) \; \label{perturb:T_i0} \;.
\end{align}

Gauss theorem works also when the region has null boundaries (see for instance \cite{Wald:1984rg}): 
\begin{equation}
\label{Gauss}
\int_\mathcal{V} \partial_\mu T^{\mu 0} \text d^4x =  \oint_{\partial \mathcal V} T^{\mu 0} n_\mu\, \text d S\; .
 \end{equation}
The only subtlety in the case of null boundaries is that one does not know how to choose the normalization of the null vector $n^\mu$ orthogonal to the surface (of course when the boundary is null this vector also lies on the surface). A related ambiguity is that there is no natural volume form on the boundary to perform the integration, since the induced metric on a null surface is degenerate. The two ambiguities compensate each other. If one chooses a 3-form $\tilde\varepsilon$ as a volume form on the null surface one needs the covector $n_\mu$ to satisfy
\be
n \wedge \tilde \varepsilon = \varepsilon \;,
\ee 
with $\varepsilon$ the volume 4-form, the one used to perform the integration in $\mathcal{V}$ in equation \eqref{Gauss}. This equation generalises the concept of orthonormal vector in the Gauss theorem and one can show that it implies eq.~\eqref{Gauss}, see \cite{Wald:1984rg}.
In our case, if one chooses 
to perform the integral over the null boundary as $\text d t \, \text d x\, \text d y$, which corresponds to a 3-form $\tilde\varepsilon_{\alpha\beta\gamma}$ which is completely antisymmetric in the variables $t, x$ and $y$, one has to normalise the orthogonal vector $n^\mu$ such that 
\be
\frac14 \varepsilon_{\alpha\beta\gamma\delta} = n_{[\alpha} \tilde\varepsilon_{\beta\gamma\delta]} \;.
\ee 
This is satisfied by the vector $n^\mu = (1,0,0,1)$ in the Minkowski coordinates $(t,x,y,z)$.


We now apply eq.~\eqref{Gauss} to the region depicted in Fig.~\ref{fig:diagram_boundary} and use eq.~\eqref{eq:emc}. There is no dependence on $x$ and $y$, so we can factor out the surface $\text d x \, \text d y$:
\begin{equation}\label{eq:energy_conservation}
\int_{\partial \mathcal V_2}  T^{00}\, \text d z - \int_{\partial \mathcal V_1}  T^{00} \,\text d z = - \int_{\partial \mathcal V_0}  \left( T^{00} - T^{z0}\right)\text d t \;.
\end{equation}
We dropped the contribution of the surface $\partial\mathcal V_0'$ since all fields vanish on this surface. The LHS is the difference in energy between $t =T$ and $t =0$, while the RHS gives the flux of energy across $\partial\mathcal V_0$. 
As shown in Appendix \ref{app:energy}, neither $\bar\gamma$ nor $\Delta\gamma$ contribute to the energy flux across $\partial \mathcal V_0$ (intuitively GWs move parallel to this surface). Conversely, since $\pi$ only depends on $u$, it does not contribute to the difference in energy.  Since the flux of energy is only due to $\pi$, which is constant on $\partial \mathcal V_0$, the flux is proportional to $T$.
We see that the dependence $\Delta\gamma \propto T$ in the LHS is necessary for the cancellation. 
Notice that the sign of $\Delta\gamma_{ij}$ in \eqref{eq:GammaBR} is the correct one: it implies that the amplitude of GW is \emph{decreasing}.
Indeed, since the total energy is just the sum of the kinetic energy of $\gamma$ and $\pi$, $\gamma$ must decrease in amplitude as $\pi$ grows.   In Appendix \ref{app:energy} we check these statements and verify eq.~\eqref{eq:energy_conservation}.

\section{Nonlinearities\label{sec:selfint}}
We now want to look at the effects of $\pi$ non-linearities (non-linearities of $\gamma$ are suppressed by further powers of $\MP$). In particular, we are going to study the effect of non-linearities on the exponential amplification of dark energy fluctuations that occurs in the narrow-resonance regime.

\subsection{$m_3^3$-operator}
\label{sec:nlt}

We start from the operator  \eqref{eq:m3term}. In this case, the  cubic self-interaction in the Lagrangian  is 
\begin{equation}\label{eq:CubicEstimate}
 \frac{1}{M^3} \Box \pi \, (\partial_i \pi)^2 \;,
\end{equation}
where $\Box \equiv - \partial_t^2 + \nabla^2 $
and
\begin{equation}
M^{3} \equiv - \frac{[ 3 m_3^6 + 4 \MP^2 (c+ 2 m_2^4)]^{3/2} }{\sqrt{2} m_3^3  \MP^3 }  = \frac{{\alpha}^{3/2}}{\alpha_{\rm B}} \left( \frac{H}{H_0} \right)^2 \Lambda_3^3\;.
\end{equation}

 At early times, when $\pi$ is in the vacuum state, it is safe to neglect these terms. However, as $\pi$ grows their  importance increases and they become comparable to the resonance term
\begin{equation}
 \frac{1}{\Lambda^2}\dot\gamma_{ij} \partial_i \pi \partial_j \pi \;.
\end{equation}
Since $M \ll \Lambda$, we expect  this to happen rather quickly.
Comparing the above operators, this takes place when $\Box \pi \sim  ({M^3}/{\Lambda^2}) \dot \gamma_{ij} \sim \sqrt{\alpha} {H} \dot \gamma_{ij}$,  which can be written as
\begin{equation}\label{eq:NLcomp}
 (\partial_i \pi)^2 \sim \alpha {c_s^2} (h_0^+)^2 \Lambda_2^4\;,
 \end{equation}
by using $\dot\gamma_{ij} \sim \omega \MP  h_0^+$ and $\Box \pi \sim (\omega/c_s) \partial_i \pi$.
When this happens, both the energy density of $\pi$ and the modification of the GW, see eq.~\eqref{eq:sol_backreact}, are  small. Indeed, using the above equation one finds
\begin{equation}
\frac{\rho_\pi}{\rho_\gamma} \sim \frac{(\partial_i \pi)^2}{(\MP \omega h_0^+)^2} \sim {\alpha}\left(\frac{c_s H_0}{\omega}\right)^2 \ll 1\;, \hspace{1cm} \frac{ \Delta \gamma}{\bar \gamma} \sim \frac{v (\partial_i \pi)^2}{\Lambda^2 \MP h_0^+} \sim { v H_0  h_0^+ \alpha_{\rm B} c_s^2} \ll 1 \;.
\end{equation}

 After this point one can no longer trust the Mathieu equation and the solutions for $\pi$ used earlier,  eq.~\eqref{eq:narrow_res_approx}: 
non-linear terms change the fundamental frequency of the oscillator in eq.~\eqref{eq:3}, so that originally unstable modes are driven out of  their instability bands and a more sophisticated analysis is required.
The same conclusion can be  obtained from the Boltzmann analysis of App.~\ref{Bose}. When the resonance term is modified by $\sim \mathcal O (1)$ corrections, particles are produced also outside the thin-shell of momenta $\Delta k$. This dispersion in momentum space implies that the Bose-enhancement factor, and in turn the growth index $\mu_k$, are affected.

The above estimate is also in agreement with numerical results in the preheating literature (see e.g.~\cite{Prokopec:1996rr,Adshead:2017xll}). These  simulations suggest that even small self-interactions of the produced fields are enough to qualitatively change the development of the resonance. 
In these works it was also shown that attractive potentials for the reheated particles modify, and eventually shut down, the exponential growth found at linear level. This conclusion is not surprising, since in these cases large field expectation values  contribute positively to their effective mass, making the decay kinetically disfavoured as soon as large field values are reached. This result cannot be applied to our case because the derivative self-interactions in eq.~\eqref{eq:CubicEstimate} do not enter with a definite sign in the action. To reach a definitive answer, in the narrow resonance regime, one would need a full numerical analysis. 

\subsection{\texorpdfstring{$\tilde m_4^2$}{tilde m42}-operator}
\label{sec:nltm4}

We will now take  $m_3^3=0$ and focus on the self-interaction of $\pi$ generated by the operator of eq.~\eqref{eq:bh}. \textcolor{black}{This choice results technically natural since $m_3^3$ corresponds, in the covariant theory, to cubic Horndeski operators that feature a weakly-broken Galilean invariance (for details, see the discussion in \cite{Pirtskhalava:2015nla, Santoni:2018rrx,Creminelli:2018xsv}).}

\textcolor{black}{Clearly, on sub-Hubble scales the most important nonlinearities are due to operators containing two derivatives per  field.}
Since we are interested in the regime $\alpha_{\rm H} \ll 1$, in this case the most relevant non-linearities are not cubic but quartic. Indeed cubic non-linearities are suppressed by $\alpha_{\rm H} /\alpha^{3/2} \cdot \partial^2\pi/\Lambda_3^3$, while the quartic ones by $\alpha_{\rm H}/\alpha^2 \cdot (\partial^2\pi/\Lambda_3^3)^2$. Thus, for $\alpha_{\rm H} \ll 1$, quartic non-linearities become relevant for a smaller value of $\partial^2\pi/\Lambda_3^3$ (unless $\alpha$ is huge, we will not be interested in this regime below). Notice that the cut-off of the theory is thus of order $\Lambda_3 \alpha^{1/3}\alpha_{\rm H}^{-1/6}$. This scale is much larger than $\omega$ for the values of $\alpha_{\rm H}$ we are interested in, so that the GW experimental results are well within the regime of validity of the theory. See Fig.~\ref{fig:EScales} for a comparison of the scales in the problem.

\begin{figure}[h]
	\centering
	\begin{tikzpicture}[xscale=7,yscale=2]
	\fill[fill=myblue!100,opacity=0.8] (0,-0.1) rectangle (.7,.5); 
	\node[above] at (0,2.8) {Energy}; 
	\draw[->,>=stealth][draw=black, very thick] (0,-.15) -- (0,2.8); 
	\draw[-][draw=black, very thick] (0,0) -- (.7,0); 
	
	\node[above] at (0.35,.1) {LIGO/Virgo}; 
	\node[right] at (.7,0) {$\textcolor{myred}{\Lambda_3} = (H_0^2\MP)^{1/3} \sim 10^{-13} ~\rm{eV}$}; 
	\draw[-][draw=black, very thick] (0,0.8) -- (.7,0.8); 
	\node[right] at (.7,0.8) {$\textcolor{myred}{\Lambda_{\rm cut-off}} \simeq \alpha_{\rm{H}}^{-1/6}\alpha^{1/3}\Lambda_3$};
	\draw[-][draw=black, very thick] (0,1.5) -- (.7,1.5); 
	\node[right] at (.7,1.5) {$\textcolor{myred}{\Lambda_\star} \simeq \alpha_{\rm{H}}^{-1/3}\alpha^{1/3}\Lambda_3$};
	\draw[-][draw=black, very thick] (0,2.2) -- (.7,2.2); 
	\node[right] at (.7,2.2) {$\textcolor{myred}{\Lambda_2} = (H_0\MP)^{1/2} \sim 10^{-3} ~\rm{eV}$};
	
	\end{tikzpicture}
	\caption{~The various energy scales in the limit $\alpha_{\rm H} \ll 1$. \textcolor{black}{We assume $\alpha \gg \alpha_{\rm H}^{1/2}$ so that $\Lambda_{\rm cut-off} \gg \Lambda_3$.  Notice that in this paper we do not consider the regime of parametrically small $c_s$, which may modify the estimates above.}} \label{fig:EScales}
\end{figure}
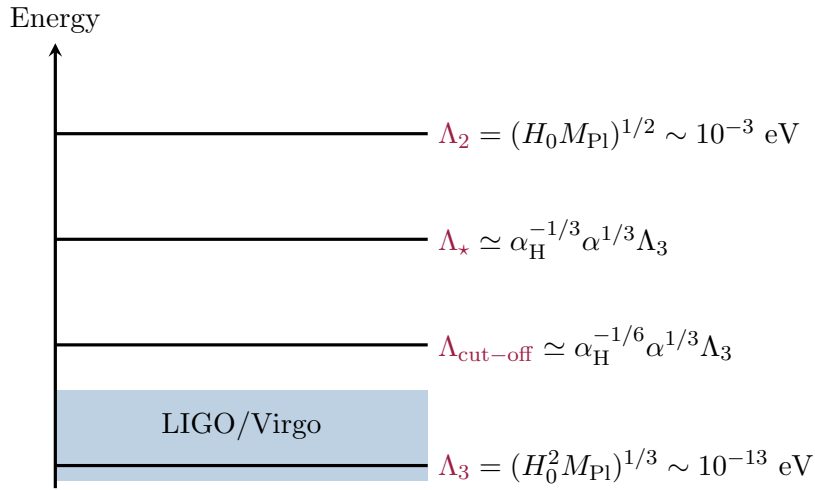

Let us start with a naive estimate for $c_s \sim1$. The resonance will be affected by non-linearities when
\be
\frac{\alpha_{\rm H}}{\alpha^2} \left(\frac{\partial^2\pi}{\Lambda_3^3}\right)^2 \sim \beta \;.
\ee
Going through the same calculations as in the case of the $m_3^3$-operator one gets the constraint
\be
\frac{ \Delta \gamma}{\bar \gamma} \sim \frac{v \,\omega (\partial_i \pi)^2}{\Lambda_\star^3 \MP h_0^+} \lesssim \beta\alpha (v H_0) \frac{H_0}{\omega h_0^+} \;.
\ee
For typical values of the parameters the RHS of this inequality is at most of order unity. This would mean that non-linearities are important when the GW signal is substantially modified. 

Actually, it turns out that the estimate above is not quite correct, as a consequence of the detailed structure of the quartic Galileon interaction. We want to evaluate the importance of the interactions on the modes that grow fastest, i.e.~on the saddle point. Using $(u,s,x,y)$ coordinates, one has $p_s =0$ on the saddle, see eqs.~\eqref{eq:ps} and \eqref{eq:solSP}. This can be understood from the equation of motion, eq.~\eqref{eq:eomp}: for a given frequency of a $\pi$ wave, i.e.~for a given momentum $|\vect p|$, one maximises the forcing term if $p_s =0$. Therefore in these coordinates the derivative with respect to $s$ vanishes. However, since the coordinate $u$ is null, the inverse metric satisfies $g^{uu} =0$. This means that also the derivative with respect to $u$ will not appear in the interaction (there are only cross terms $\propto \partial_u\partial_s$). Therefore we are left only with the two coordinates $x$ and $y$. However, the structure of the quartic Galileon is such that it vanishes if one has less than three dimensions. We conclude that the quartic $\pi$ self-interaction vanishes on the saddle point. To get a correct estimate one is forced to look at deviations from the saddle, i.e.~one has to estimate what is the typical range of $p_s$ around $p_s =0$ that contributes to the integrals like eq.~\eqref{eq:J1}. 

The rms value of the various variables can be read from the inverse of the Hessian matrix eq.~\eqref{eq:matrixH}. 
The momentum $p_s$ can be written in terms of the integration variables $\{\xi, \Omega, \varphi\}$ as
\be
 p_s = \frac{c_s \omega}{1-c_s^2}\sqrt{\xi} (\Omega-c_s) \;.
\ee
One can write
\begin{equation}\label{eq:uncertPs}
(\Delta p_s)^2 = \frac{\partial p_s}{\partial X^i} \frac{\partial p_s}{\partial X^j}\Delta X^i \Delta X^j = \frac{\partial p_s}{\partial X^i} \frac{\partial p_s}{\partial X^j} (\text H^{-1})^{ij}\;.
\end{equation}
Evaluated at the saddle, the matrix of the first derivatives of $p_s$ is zero, except for the entry $(\Omega, \Omega)$ which is given by
\begin{equation}\label{eq:DerivPs}
\left(\frac{\partial p_s}{\partial \Omega}\right)^2  = \frac{c_s^2 \omega^2}{(1-c_s^2)^2(4- \beta^2)} \;.
\end{equation}
The entry $(\Omega, \Omega)$ of the inverse of the Hessian reads $\left(\text H^{-1}\right)^{\Omega \Omega} = \frac{(1-c_s^2)^2}{4 \beta \tau} (4- \beta^2)^{3/2}$\;.
Therefore we obtain
\begin{equation}\label{eq:EstimPs}
\Delta p_s = \frac{c_s \omega (4- \beta^2)^{1/4}}{2 \sqrt{\beta \tau}} \simeq \frac{c_s \omega}{\sqrt{2 \beta \tau}}\;.
\end{equation}
Using this estimate one can revise the bound above as
\be\label{bound}
\frac{ \Delta \gamma}{\bar \gamma} \sim \frac{v \,\omega (\partial_i \pi)^2}{\Lambda_\star^3 \MP h_0^+} \lesssim \beta c_s^3 \alpha (v H_0) \frac{H_0}{\omega h_0^+} \sqrt{\beta \tau}\;.
\ee

It is important to stress that the coefficient of the quartic self-interaction of $\pi$ is tied by symmetry with the one of the operator $\ddot\gamma_{ij} \partial_i\pi \partial_j\pi$, so that the effect of self-interactions cannot be suppressed. This statement holds even considering models with $c_{\rm T} \neq 1$: since we are considering a regime of very small $\alpha_{\rm H}$, comparable values for $c_{\rm T}-1$ are not ruled out experimentally. If one does not impose the constraints on the speed of GWs, instead of the single operator of eq.~\eqref{eq:bh} one has two independent coefficients
\be
\frac{\tilde{m}_4^2}{2} \delta g^{00}\, {}^{(3)}\!R + \frac{{m}_5^2}{2} \delta g^{00} \left(\delta K_\mu^\nu \delta K_\nu ^\mu  - \delta K^2 \right) \;.
\ee
In terms of these parameters, the coefficient of the quartic Galileon was calculated in \cite{Crisostomi:2019yfo} and it reads
	\be
	\frac{2}{\MP^2}(\tilde{m}_4^2 + {m}_5^2) \;.
	\ee
	On the other hand the coefficient of the operator $\ddot\gamma_{ij} \partial_i\pi \partial_j\pi$ reads (see App.~B of \cite{Creminelli:2018xsv})
	\be
	\frac{1}{\MP^2}(\tilde{m}_4^2 + {m}_5^2 c_{\rm T}^2) \;.
	\ee
	The two coincide, modulo a factor of 2, up to relative corrections suppressed by $c_{\rm T}^2-1$.

\section{Observational signatures for \texorpdfstring{$\tilde m_4^2$}{tilde m42}}
\label{sec:mtilde4}

\subsection{Fundamental frequency}
\label{sec:mtilde41}

In the narrow resonance regime, $\beta \ll 1$, with the replacement $\Lambda^2 \rightarrow \Lambda_\star^3 \omega^{-1}$, from eq.~\eqref{eq:GammaBR} one gets a relative modification of the GW given by \textcolor{black}{
	\begin{equation}\label{eq:GammaBRm4}
	\frac{\Delta \gamma(u, v)}{\bar \gamma} \simeq -\frac{v}{4 \Lambda_\star^3 \MP h_0^+}  \frac{  (1-c_s^2)^2}{  c_s^5  \sqrt{ \beta}} \,   \frac{\omega^{7/2}}{(8  u \pi )^{3/2}} \exp \left( \frac{\beta}{4} \omega u \right)  \;.
	\end{equation}}


The amplitude of GWs in the plus polarization is given by \cite{Maggiore:1900zz}
\be
\label{numbcyc}
h_0^+ \sim \frac4{r} (G M_c)^{5/3} (\pi f)^{2/3} \;,
\ee
where $r$ is the distance from the source and $M_c$ is the chirp mass of the binary system. 
The number of cycles of the gravitational wave is given by 
\be
N_{\rm cyc} = \frac{\omega u}{2 \pi} = f u \;, 
\ee
where we have defined the gravitational wave frequency $f \equiv \omega /2 \pi$. In our calculations we took a time-independent frequency, while in reality $f$ increases with time, during the binary inspiralling. The frequency can be taken as roughly constant for the number of cycles $N_{\rm cyc}$ required for $f$ to double in size.
In particular, the number of cycles between $f$ and $2 f$ is given by \cite{Maggiore:1900zz} \textcolor{black}{
	\be\label{eq:cyc}
	\bar N_{\rm cyc} = \frac{4 - 2^{1/3}}{128 \pi } (\pi GM_c f)^{-5/3} \;.
	\ee
}

Inverting these relations we can express the chirp mass as a function of the number of cycles and the frequency. Using this in eq.~\eqref{numbcyc} we obtain \textcolor{black}{
	\be\label{h0N}
	h_0^+ \sim \frac{0.0087}{f \bar N_{\rm cyc} r} \;.
	\ee
}
In the calculation we approximate the GW amplitude as a constant, therefore we have to limit $v \lesssim r$.

Expressing $\beta$ in term of $\alpha_{\rm H}$, replacing $h_0^+$ from the above relation and using  $v = r$, eq.~\eqref{eq:GammaBRm4} becomes
\be\label{eq:expo}
\frac{\Delta \gamma}{\bar \gamma} \sim - 0.006\times \frac{ (1-c_s^2)^2}{c_s^3} \left( \frac{\alpha_{\rm H}}{\alpha c_s^2} \right)^{\frac12}   \left( \frac{H_0}{\MP} \right)^{\frac16} \left( \frac{2 \pi f}{\Lambda_3} \right)^{\frac{11}2} ( r H_0)^{\frac52} \exp \left[  \frac{0.12}{ r H_0} \, \left( \frac{\MP}{H_0} \right)^{\frac13}  \frac{\alpha_{\rm H}}{\alpha c_s^2} \, \frac{2 \pi f}{\Lambda_3}  \right] \;.
\ee
Note that this expression is independent of \textcolor{black}{$\bar N_{\rm cyc}$}.
Sizeable effects in the GW waveform can be obtained when the argument of the exponential is $\sim {\cal O} (10^2)$, which translates into
\be
\alpha_{\rm H}   \gtrsim 10^{- 17} \cdot  r H_0 \cdot \frac{\Lambda_3}{2 \pi f} \alpha c_s^2\;.
\ee
We have also to impose the constraint of eq.~\eqref{bound}, which using eq.~\eqref{h0N} reads 
	\be
	\frac{ \Delta \gamma}{\bar \gamma} \lesssim 18 (\beta \bar N_{\rm cyc})^{3/2} c_s^3 \alpha (r H_0)^2  \equiv \left(\frac{\Delta \gamma}{\bar \gamma}\right)_{\textrm{NL}} \;,
	\ee
	where the term $\beta \bar N_{\rm cyc}$ roughly coincides with the argument of the exponential in eqs.~\eqref{eq:GammaBRm4} and \eqref{eq:expo}. Finally, a further constraint to impose is the narrow resonance condition $\beta \ll 1$.

\begin{figure}[t]
	\begin{subfigure}{0.5\textwidth}
		\includegraphics[width=\linewidth]{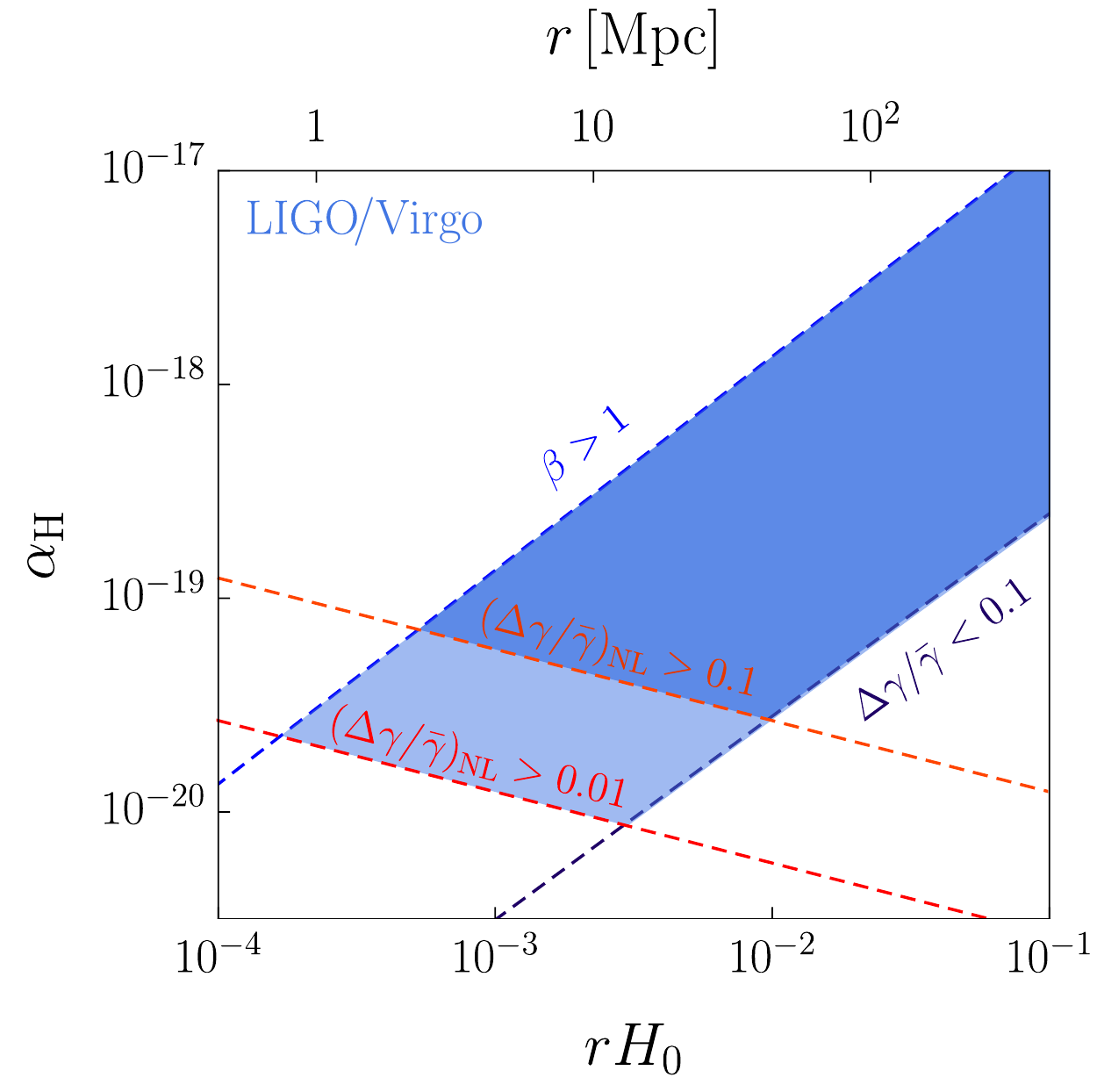}
		\caption{\label{fig:LIGO}} 
	\end{subfigure}
	\hspace*{\fill} 
	\begin{subfigure}{0.5\textwidth}
		\includegraphics[width=\linewidth]{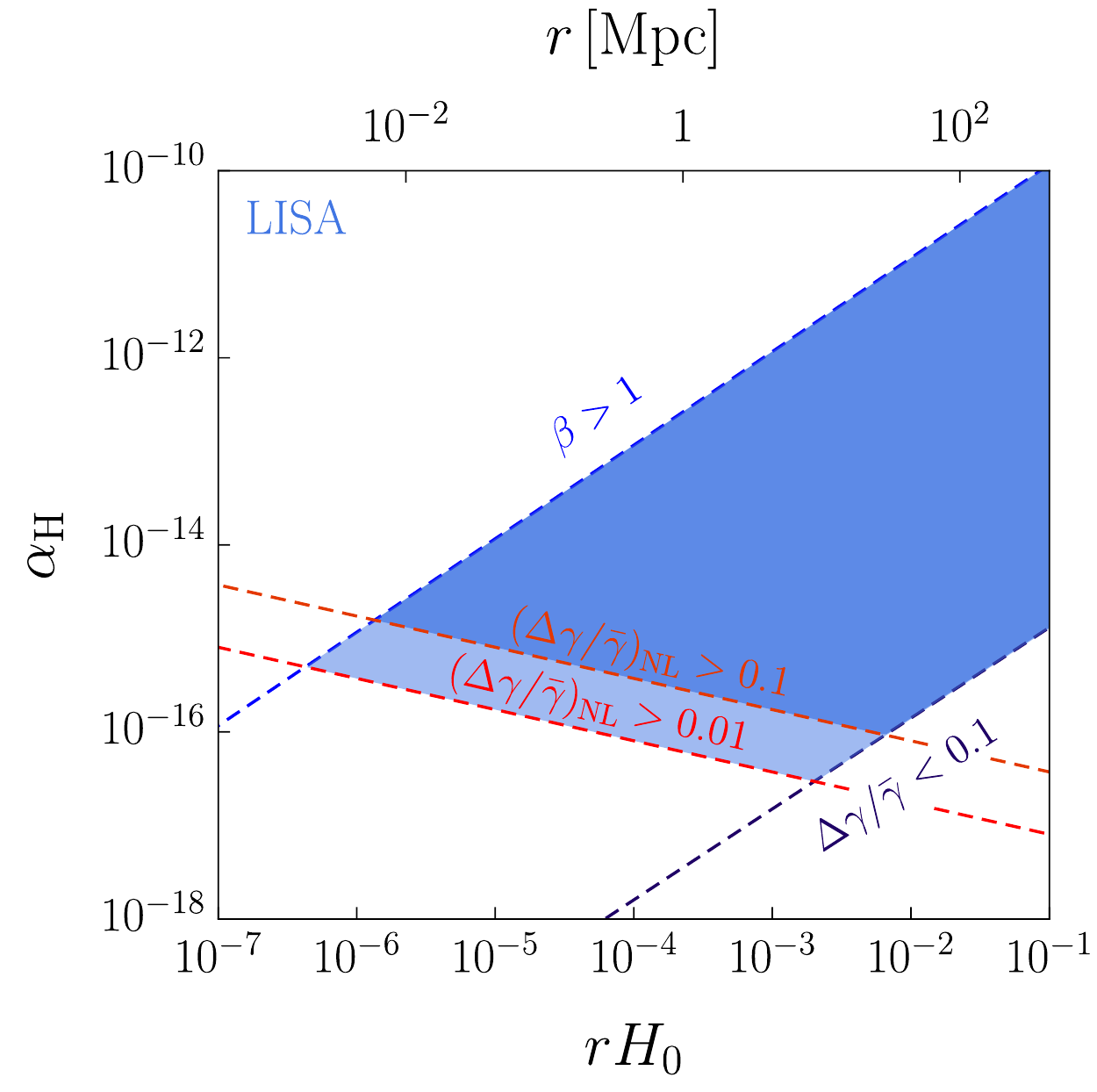}
		\caption{\label{fig:LISA}} 
	\end{subfigure}
	\hspace*{\fill} 
	\caption{~The blue regions indicate where our approximations apply and one has a sizeable modification of the GW signal. Above the red dashed lines the effect of $\pi$ non-linearities is small. The upper blue dashed lines indicate the line $\beta =1$: our analytical approximation holds for $\beta \ll 1$ \textcolor{black}{and that we extrapolate to $\beta \lesssim 1$}. One has a sizeable effect on the GW above the lower dashed blue lines.  This mainly depends on the exponential of \textcolor{black}{ $\beta \bar N_{\rm cyc}$ and since $\bar N_{\rm cyc}$} is fixed once $M_c$ and $f$ are given, this constraint is to a good approximation a lower bound on $\beta$. 
		Left panel (Fig.~\ref{fig:LIGO}): LIGO/Virgo case: $f = 30$~Hz, $M_c = 1.188~ \rm M_\odot$ as for GW170817. Right panel (Fig.~\ref{fig:LISA}): LISA case: $f = 10^{-2}$~Hz, $M_c = 30~\rm M_\odot$ as for GW150914.} 
\end{figure}
In Figs.~\ref{fig:LIGO} and \ref{fig:LISA} we plot these three constraints as a function of $\alpha_{\rm H}$ and the distance between the source and the resonant decay of the GW into $\pi$ (therefore this is not the distance between the source and the detector). The first plot is done for a ground-based interferometer with LIGO/Virgo-like sensitivity; in order to maximise the number of oscillation we choose a neutron-star event similar to GW170817 \cite{TheLIGOScientific:2017qsa}. The second plot is for a space-based interferometer with LISA-like sensitivity \cite{Audley:2017drz} and a binary black hole event similar to GW150914 \cite{Abbott:2016blz}. The blue region corresponds to a sizeable modification of the GW signal, calculable within our approximation. The neutron-star merger GW170817 is at a distance of approximately 40 Mpc. The absence of sizeable effects ($\Delta \gamma > 0.1 \gamma$\textcolor{black}{; of course future measurements will improve this sensitivity}) in the observed event puts constraints on a resonant effect that takes place at less than 40 Mpc from the source and rules out the interval: \textcolor{black}{$3 \times 10^{-20} \lesssim \alpha_{\rm H} \lesssim 10^{-18}$. Future measurements by LISA of events similar to GW150914 can, on the other hand, constrain the range $10^{-16} \lesssim \alpha_{\rm H} \lesssim 10^{-10}$.}

\textcolor{black}{It may be useful to summarize here the set of assumptions for the validity of the constraints in these figures:} 
\begin{itemize}

\item Narrow resonance regime, i.e.~$\beta \sim \alpha_{\rm H} (\omega /H)^2 h_0^+ \lesssim 1$. This is diplayed by the blue-dashed line in the figures.

\item The modification of the GW  is sufficiently sizable to be observed, i.e.~$|\Delta \gamma| > 0.1  \gamma$. This is diplayed by the purple-dashed line in the figures.
The modification  takes place outside the Vainshtein radius, which for the values of $\alpha_{\rm H}$ that we are considering corresponds to distances from the source smaller than those shown on the horizontal axis. 
Moreover, to be relevant the modification must occur before the detection.

\item Nonlinearities in $\pi$ are small so that their effect on the Mathieu equation is negligible, see discussion in Sec.~\ref{sec:nltm4}. This is displayed by the red-dashed lines in the figures.

\item In the figures we assume $c_s = 1/2$ and the constraints would change for different values of $c_s$. From eq.~\eqref{eq:GammaBR} the effect is proportional to $(1-c_s^2)^2$ so that it is suppressed for values of $c_s$  close to 1. Moreover, as we discussed in Sec.~\ref{subsec:ParametricResonance}, for $c_s -1\lesssim 10^{-14}$ the $\pi$ lightcone is too wide and our calculations do not apply.   
(Notice also that in our estimates of the various scales we are assuming that $c_s$ is not parametrically small.)

\end{itemize}

For comparison, the bound coming from the perturbative decay of the graviton, see eq.~(46) of \cite{Creminelli:2018xsv}, reads $\alpha_{\rm H} \lesssim 10^{-10}$. Such values of $\alpha_{\rm H}$ are outside the narrow resonance regime studied in this paper. One expects that the effect of large occupation number and nonlinearities of $\pi$  are also important for larger values of $\alpha_{\rm H}$ so that this perturbative bound should  be revised. 
Moreover, notice that when the GW is closer to the source, its amplitude is larger and $\beta$ will exceed unity. At a certain point one enters the Vainshtein regime and the coupling $\gamma\pi\pi$ is reduced (since we are considering very small values of the coupling, the Vainshtein radius will be correspondingly small). We will study all these aspects in a forthcoming publication \cite{Creminelli:2019new}.

\subsection{Higher harmonics and precursors}
\label{sec:mtilde42}
Since for the operator $\tilde m_4^2$ one can trust the parametric growth and the change in the GW signal $\Delta\gamma$, we want to study this correction in more detail.  At leading order in $q$ the change $\Delta\gamma$ has the same frequency as the original wave. However, corrections to the leading solution \eqref{eq:narrow_res_approx} introduce higher harmonics in the GW signal.
Higher harmonics appear in two quantitatively different ways. 
First, we can consider higher instability bands, see Fig.~\ref{fig:mathsigmas}. However, in the narrow resonance approximation the instability coefficient in the $m$-th band scales as $\mu_m \sim q^m / (m!)^2 \sim \beta^m$ \cite{mclachlan1951theory,olver2010nist}. Therefore, in the long $\tau$ limit they are exponentially suppressed compared to the fundamental band.

Second, we can stay in the first band and look at the modification of the $\pi$ solution at higher orders in $q$. This will give an effect which is only suppressed by positive powers of $q$. If we include the first two corrections to eq.~\eqref{eq:narrow_res_approx}, we have\footnote{In this section we write the formulas for the case of $m_3^3$, so that they can be easily compared with Section \ref{subsec:waveformmod}. Notice that, since in this case $\pi$ is coupled with $\dot\gamma$, while in the case $\tilde m_4^2$ it is coupled with $\ddot\gamma$, there is an overall shift of $\pi/2$ between the two cases.}
\be
\begin{split}
	f(\tau) &\simeq c_+ e^{(\mu + \delta \mu)\tau} F(\tau; \sigma) + c_- e^{(-\mu - \delta \mu)\tau} F(\tau;-\sigma)  \\ 
	F(\tau; \sigma) &= \sin(\tau - \sigma) + s_ 3 \sin(3 \tau - \sigma) + c_ 3 \cos(3 \tau - \sigma) + s_ 5 \sin(5 \tau - \sigma) \\
	s_ 3 &\equiv -\frac{1}{8}q+ \frac{1}{64}q^2 \cos(2 \sigma)\;, \hspace{0.4cm}
	c_ 3 \equiv   \frac{3}{64}q^2 \sin(2 \sigma) \;, \hspace{0.4cm}s_ 5 \equiv  \frac{1}{192}q^2\; .
\end{split}
\ee
Clearly, in this case the higher frequency component grows as fast as the original frequency. Therefore it is not necessary to specify the correction $\delta \mu$.
The correction to $|f_{\tilde{p}}|^2$ at order $q^2$ is thus
\be
\begin{split}
	|f_{\tilde{p}}|^2 \simeq |c_+|^2 e^{2 \mu \tau}&\left[\sin^2(\tau-\sigma) +2 s_3 \sin(\tau-\sigma)\sin(3\tau-\sigma) + s_3^2\sin^2(3\tau-\sigma) + \right. \\
	& \left. \quad  2c_3\sin(\tau-\sigma)\cos(3\tau-\sigma) + 2 s_5 \sin(\tau-\sigma)\sin(5\tau-\sigma) \right] \;.
\end{split}
\ee

On the other hand, when $q$ becomes negative the solution can be obtained by performing the transformation $\tau \rightarrow \tau + \pi / 2$, or equivalently $q\rightarrow-q$ together with $\sigma \rightarrow -\sigma - \pi / 2$.
By using the latter transformation and by recalling that by flipping the sign of $q$ we send the decreasing mode into the growing one, we get the contribution for negative $q$:
\be
\begin{split}
	|f_{\tilde{p}}|^2 \simeq |c_+'|^2 e^{2 \mu \tau}&\left[\cos^2(\tau-\sigma) -2 s_3 \cos(\tau-\sigma)\cos(3\tau-\sigma) + s_3^2\cos^2(3\tau-\sigma) + \right. \\
	& \left. \quad 2c_3\cos(\tau-\sigma)\sin(3\tau-\sigma) + 2 s_5 \cos(\tau-\sigma)\cos(5\tau-\sigma) \right]\; .
\end{split}
\ee
By direct calculation we can also show that, as in \eqref{eq:SPscc}, $|c_+| = |c_+'| = 1$ at lowest order in $q$.
We also note that, since all the frequencies in the expressions above are multiples of $\omega$, the position of the saddle point is not affected and it is simply fixed by the exponential function. Hence the expression for $J(u)$ can be obtained by using the saddle-point procedure as in \eqref{eq:fullJ}. We obtain
\be
\begin{split}\label{eq:GammaBRhigher}
	\Delta \gamma_{ij} &(u, v)  \simeq -\frac{v}{4 \Lambda^2}  \frac{  (1-c_s^2)^2}{  c_s^5  \sqrt{ \beta}} \,   \frac{\omega^{5/2}}{(8  u \pi )^{3/2}}    \exp \left( \frac{\beta}{4} \omega u \right) \epsilon_{ij}^+  \\
	& \cdot \left\{\sin \left( \omega u + \frac{\beta}{2} \right) -\frac{\beta}{8} \cos(\omega u)  + \frac{\beta^ 2}{768}  \left[ 12 \sin\left(\omega u - \frac{\beta}{2}\right) + 6 \sin\left(\omega u + \frac{\beta}{2}\right) + \sin\left(3 \omega u + \frac{\beta}{2}\right) \right] \right\} \;.
\end{split}
\ee
(The result for the case of $\tilde m_4^2$, which is the focus of this section, can be obtained by the replacement $\Lambda^2 \to \Lambda_\star^3 \omega^{-1}$.) We are interested in the $\tau$-dependent oscillatory part of \eqref{eq:GammaBRhigher}. Compared to \eqref{eq:fullJ} we get a correction at the fundamental frequency linear in $\beta$ (with no phase-shift). Moreover, to find the first higher harmonics we have to go to order $\beta^2$, where indeed we encounter a term with three-times the frequency $\omega$. Note that by going to the second band instead, we would find higher harmonics of frequency $2 \omega$. 

Higher harmonics enter in the bandwidth of the detector at earlier time compared to the main signal. In some sense they are {\em precursors} of the main signal. To assess their potential observability, one has to consider that also the post-Newtonian expansion generates terms with a frequency which is a multiple of the original one. We leave this study to future work.

\section{Conclusions and future directions}
\label{sec:conclusion}

We have studied the effects of a classical GW on scalar field fluctuations, in the context of dark energy models parametrized by the EFT of Dark Energy. 
The GW acts as a classical background and modifies the dynamics of dark energy perturbations, described by $\pi$, leading to parametric resonant production of $\pi$ fluctuations. 
This regime is described by a Mathieau equation, where modes in the unstable bands grow exponentially. In the regime of narrow resonance, corresponding to a small amplitude of the GW, the instability can be studied analytically in detail. One can also include the back-reaction of the produced $\pi$ on the original GW, which leads to a change in the signal. The resonant growth is however very sensitive to the non-linearities of the produced fields: when non-linearities of $\pi$ are sizeable the resonance is damped. This regime cannot be captured analytically: numerical simulations are needed to confirm this effect. This happens for the operator $m_3^3$, while a sizeable modification of the GW signal is possible
for the operator $\tilde m_4^2$, at least in some range of parameters, see Figs.~\ref{fig:LIGO} and \ref{fig:LISA}.

Many questions remain open and will be the subject of further studies. For larger amplitudes of the GW, one exits the regime of narrow resonance: the GW may induce instabilities, which will be studied in \cite{Creminelli:2019new}. This will also allow to revise the robustness of the perturbative calculation of \cite{Creminelli:2018xsv}. In the same paper we will also study how these effects are changed in the presence of Vainshtein screening.  
Simulations are probably required if one wants to study the regime of parametric resonance in the presence of sizeable $\pi$ self-interactions. Similar effects are at play in preheating after inflation and are also addressed numerically, see e.g.~\cite{Prokopec:1996rr,Adshead:2017xll}. It would also be interesting to try to envisage methods to look for the small changes in the GW signal. For instance, the generation of higher harmonics that enter the detection bandwidth before the main signal is a striking possible effect if it can be disentangled from post-Newtonian corrections. The effects that we studied look quite generic to all theories in which gravity is modified with a quite low cut-off. Therefore, it would be nice to explore other setups, starting from the DGP model. This model features the same non-linear interaction of $\pi$ as the ones of the $m_3^3$ operator; however the extra-dimensional origin changes the dynamics of GWs so our analysis cannot be applied directly. We hope to come back to all these points in the future.

\section*{Acknowledgements}
It is a pleasure to thank M.~Celoria, G.~Cusin, P.~Ferreira, M.~Lagos, M.~Lewandowski, S.~Melville, A.~Nicolis, and F.~Schmidt for useful discussions. We also thank the anonymous referee for a very careful reading of the article and insightful comments. P.~C.~and F.~V.~acknowledge the kind hospitality of PUCV, Valparaiso, Chile where part of this work was carried on.

\appendix
\section{Interactions in Spatially-Flat Gauge\label{app:spatflat}}
In this appendix we are interested in redoing the computation of Section \ref{subsec2.1} in the spatial-flat gauge. Here the metric is written in the general decomposition (see e.g. \cite{Malik:2008im})
\begin{equation}
\textrm ds^2 = -(1+\delta N)^2 \textrm d\tilde{t}^2 + a(\tilde{t})^2(e^{\tilde{\gamma}})_{ij}(\textrm d\tilde{x}^i + \tilde{N}^i \textrm d\tilde{t})(\textrm d\tilde{x}^j + \tilde{N}^j \textrm d\tilde{t}) \label{SpatiallyFlatMetric}
\end{equation}
with $\tilde{\gamma}_{ij}$ being transverse, $\tilde{\partial}_i\tilde{\gamma}_{ij} = 0$, and traceless $\delta_{ij}\tilde{\gamma}_{ij} = 0$. The shift vectors $\tilde{N}_i$ can be decomposed as $\tilde{N}_i = \tilde{\partial}_i\tilde{\psi} + \hat{\tilde{N}}_i$ where $\tilde{\partial}_i\hat{\tilde{N}}_i = 0$ ($\tilde{\partial}_\mu \equiv \partial/\partial \tilde{x}^\mu$). Note that in this gauge $\tilde{\pi}(\tilde{x})$ denotes the Goldstone field. Since both $\delta N$ and $\tilde{N}_i$ are  Lagrange multipliers in this spatially-flat gauge, by solving the constraint equations (see e.g. \cite{Creminelli:2018xsv,Maldacena:2002vr}) they can be expressed up to first order as
\begin{equation}
\delta N = \frac{m_3^3}{m_3^3-2\MP^2H}\dot{\tilde{\pi}} + \frac{2\MP^2\dot{H}}{m_3^3-2\MP^2H}\tilde{\pi} \equiv \alpha_N\dot{\tilde{\pi}} + \tilde{\alpha}_N\tilde{\pi} \label{DeltaN}
\end{equation}
and 
\begin{equation}
\tilde{\psi} = -\frac{m_3^3+4H\tilde{m}_4^2}{m_3^3-2H\MP^2}\tilde{\pi} - \frac{3m_3^6H - 4H\MP^2(\dot{H}\MP^2 - 2m_2^4)}{(m_3^3-2\MP^2H)^2}\frac{a^2}{\tilde{\nabla}^2}\dot{\tilde{\pi}} \equiv \alpha_\psi\tilde{\pi} + \tilde{\alpha}_\psi\frac{a^2}{\tilde{\nabla}^2}\dot{\tilde{\pi}}\; , \label{Psi}
\end{equation}
where $\tilde{\nabla}^2 \equiv \tilde{\partial}_i\tilde{\partial}_i$. Notice that we have kept the non-local term in (\ref{Psi}) since this will contribute to the vertex that we are going to consider. According to \cite{Creminelli:2018xsv}, $\tilde{\pi}$ is canonically normalised as
\begin{equation}
\tilde{\pi} = \frac{2H\MP^2-m_3^3}{\sqrt{2}H\MP[3m_3^6+4\MP^2(c+2m_2^4)]^{1/2}}\pi_c \; , \label{canoPi}
\end{equation}
while the canonical normalisation of $\tilde{\gamma}_{ij}$ remains the same as in the Newtonian gauge. Here we keep a distinction between the canonical field $\pi_c$ and the non-canonical field $\tilde{\pi}$ unlike in the main text.

We are going to investigate the non-linear $\gamma\pi\pi$ vertex which contains three or more derivatives (we will drop all the rest). In principle, this vertex gets contributions from the action (\ref{starting_action}) through the Einstein-Hilbert and the $S_{m_3}$ terms. Let us first consider the contribution from $S_{m_3}$ term. As usual, the extrinsic curvature is defined by 
\begin{equation}
\tilde{K}_{ij} = \frac{1}{2N}(\dot{\tilde{h}}_{ij}-\tilde{D}_i\tilde{N}_j-\tilde{D}_i\tilde{N}_j) \; ,
\end{equation}
where $\tilde{D}_i$ denotes the 3d covariant derivative with respect to the induced metric $\tilde{h}_{ij}$. 
It is straightforward to show that a variation of $\tilde{K}$ ($\tilde{K} = \tilde{h}^{ij}\tilde{K}_{ij}$) subjected to the metric (\ref{SpatiallyFlatMetric}) contains $a^{-2}\tilde{\gamma}_{ij}\tilde{\partial}_i\tilde{\partial}_i\tilde{\psi}$. After performing the Stuekelberg trick of eq.  (\ref{eq:Stuekelberg1}) and (\ref{eq:Stuekelberg2}), the contribution from $S_{m_3}$ which has three derivatives reads
\begin{equation}
S_{m_3} \supset \frac{2m_3^3\MP^4H^2}{(m_3^3-2\MP^2H)^2}\int \text d^4
\tilde{x}~a~\dot{\tilde{\gamma}}_{ij}\tilde{\partial}_i\tilde{\pi}\tilde{\partial}_j\tilde{\pi} \; , 
\end{equation}
where we have taken the term $-2(1-\alpha_N)\dot{\tilde{\pi}}$ from $\delta \tilde{g}^{00}$ and used (\ref{DeltaN})-(\ref{Psi}).

Using the canonical normalisations both for $\tilde{\pi}$ (\ref{canoPi}) and $\tilde{\gamma}_{ij}$  one obtains the same interaction that we have obtained in the Newtonian gauge (\ref{InterationNew}). Thus we expect that the contribution arising from $S_{EH}$ cancels out. To show this, notice that the contribution from the Einstein-Hilbert term comes from $N(\tilde{K}_{ij}\tilde{K}^{ij} - \tilde{K}^2)$. More specifically, we have    
\begin{equation}
S_{\text{EH}} \supset \frac{\MP^2}{2} \int \text d^4
\tilde{x}~a~\left(4H\delta N\tilde{\gamma}_{ij}\tilde{\partial}_i\tilde{\partial}_j\tilde{\psi} + \delta N\dot{\tilde{\gamma}}_{ij}\tilde{\partial}_i\tilde{\partial}_j\tilde{\psi} + \frac{1}{2a^2}\tilde{\nabla}^2\tilde{\gamma}_{ij}\tilde{\partial}_i\tilde{\psi}\tilde{\partial}_j\tilde{\psi}\right) \; , \label{EH}
\end{equation}
where we have performed a few integration by parts. Using (\ref{DeltaN}) and (\ref{Psi}) in $S_{\text{EH}}$ one obtains
 \begin{align}
 S_{\text{EH}} &\supset \frac{\MP^2}{2} \int \text d^4
 \tilde{x}~a~\left\{\alpha_\psi(\alpha_N+\alpha_\psi)\dot{\tilde{\pi}}\dot{\tilde{\gamma}}_{ij}\tilde{\partial}_i\tilde{\partial}_j\tilde{\pi} - a^2\tilde{\alpha}_\psi(\alpha_N + \alpha_\psi)\tilde{\pi}\ddot{\tilde{\gamma}}_{ij}\tilde{\partial}_i\tilde{\partial}_j\frac{1}{\tilde{\nabla^2}}\dot{\tilde{\pi}} \right. \nonumber \\
 &+ \left. [H\alpha_\psi(2\alpha_N+\alpha_\psi) - \alpha_\psi(\tilde{\alpha}_N + \dot{\alpha}_\psi) + \alpha_N\tilde{\alpha}_\psi c_s^2]\dot{\tilde{\gamma}}_{ij}\tilde{\partial}_i\tilde{\pi}\tilde{\partial}_j\tilde{\pi} \right\} \; , \label{EH2}
 \end{align}
where we have used the linear equations of motion for $\tilde{\gamma}_{ij}$, $\ddot{\tilde{\gamma}}_{ij} + 3H\dot{\tilde{\gamma}}_{ij} -\frac{1}{a^2}\tilde{\nabla}^2\tilde{\gamma}_{ij}=0$, and for $\tilde{\pi}$, $\ddot{\tilde{\pi}} + 3H\dot{\tilde{\pi}} - \frac{c_s^2}{a^2}\tilde{\nabla}^2\tilde{\pi} = 0$.

Notice that the first two terms on RHS vanish due to the fact that $\alpha_N + \alpha_\psi = 0$. Let us consider the prefactor of the last term. Using the expression of $c_s^2$ (\ref{Cs}) and the definitions of all those $\alpha$-parameters (\ref{DeltaN})-(\ref{Psi}), the prefactor can be rewritten as 
\begin{align}
H\alpha_\psi(2\alpha_N+\alpha_\psi) - \alpha_\psi(\tilde{\alpha}_N + \dot{\alpha}_\psi) + \alpha_N\tilde{\alpha}_\psi c_s^2 = H\alpha_\psi (\alpha_N + \alpha_\psi) + \frac{H\tilde{\alpha}_N^2}{M_{Pl}^2\dot{H}^2}(M_{Pl}^2\dot{H}\alpha_\psi - c\alpha_\psi) \; .
\end{align}
In our case the coupling with matter has been neglected ($\rho_m = P_m = 0$), therefore the parameter c is equal to $-M^2_{Pl}\dot{H}$  (see e.g. \cite{Cheung:2007st,Gubitosi:2012hu}). The prefactor is then given by
\begin{align}
H\alpha_\psi(2\alpha_N+\alpha_\psi) - \alpha_\psi(\tilde{\alpha}_N + \dot{\alpha}_\psi) + \alpha_N\tilde{\alpha}_\psi c_s^2 = (\alpha_\psi + \alpha_N)(H\alpha_\psi + \frac{H\tilde{\alpha}_N^2}{\dot{H}}) 
\end{align}
which is again zero since $\alpha_N + \alpha_\psi = 0$ in this case.
Therefore $S_{EH}$ gives no contribution to our $\gamma\pi\pi$ vertex and $S_{m_3}$, on the other hand, gives the same result that we have obtained in the Newtonian gauge (\ref{InterationNew}).

\section{Parametric resonance as Bose enhancement}
\label{Bose}

In this appendix we want to reinterpret the exponential growth due to parametric resonance as the Bose enhancement of the perturbative decay $\gamma\rightarrow \pi \pi$. To see this, we study the Boltzmann equation for the number density of dark energy fluctuations. We denote by $n_{\vect{k}}^\pi$ and $n_{\vect{k}}^\gamma$  the occupation numbers, respectively, of $\pi$ and $\gamma$. Moreover, the number density for the particle species $\pi$ is defined as
\begin{equation}
    n_\pi \equiv \int \frac{\text d^3 \vect k}{(2 \pi)^3} n_{\vect{k}}^\pi \;,
\end{equation}
and an analogous definition holds for $\gamma$.

Let us consider a collection of gravitons with frequency $\omega$, each of them decaying into two $\pi$-particles. For concreteness, we will focus on the case $c_s^2 \ll 1$, for which the two momenta of $\pi$, $\vect{k}$ and $-\vect{k}$, have opposite directions and equal magnitudes $k = \omega/(2c_s)$.
Following \cite{Mukhanov:2005sc} and denoting by $\Gamma_{\gamma\rightarrow\pi\pi}$ the tree-level decay rate (see e.g.~\eqref{DecatRate}), the rate of change of $n_\pi$ in a given volume $V$ is
\be
\frac{\textrm d n_\pi}{\textrm d u} \simeq 2 \frac{\Gamma_{\gamma\rightarrow\pi\pi}}{V } [(n_{\vect{k}}^\pi+1)(n_{-\vect{k}}^\pi+1)n_{\omega}^\gamma - n_{\vect{k}}^\pi n_{-\vect{k}}^\pi (n_{\omega}^\gamma+1)] \;,
\ee
where  the factor of 2 accounts for two identical particles in the final state. As explained earlier in Sec.~\ref{subsec:ParametricResonance}, for small $c_s$ we can use $u$ as time.
On the right-hand side, we have neglected integration over the angle $\varphi$ which would appear when considering only one of the GW polarizations. For $n_{\vect{k}}^\pi=n_{-\vect{k}}^\pi=n_k^\pi$ and $n_{\omega}^\gamma \gg \{n_{\vect{k}}^\pi,1\}$ 
 we find
\be
\frac{\textrm d n_\pi}{\textrm d u} \simeq 2 {\Gamma_{\gamma\rightarrow\pi\pi}} n_\gamma  \left( 1+2 n_k^\pi \right) \ , \label{RateOfChange}
\ee
where we have introduced the number density of gravitons, here given by $n_\gamma = n_{\omega}^\gamma/V$.

The produced $\pi$-particles end up populating the spherical shell $k = k_0 \pm \Delta k/2$ of radius $k_0 \simeq \omega/(2c_s)$ and thickness $\Delta k$, 
so that their occupation number is related to their number density by
\be
n_{k}^\pi  \simeq \frac{n_\pi}{4\pi k_0^2\Delta k/(2\pi)^3} \label{Occupation} \;.
\ee
The thickness is given by comparing 
the time-independent part of the equation of motion for $\pi$, eq.~\eqref{eq:1},  with the amplitude of its  periodic part. Using $\Omega = c_s \ll1$ valid for small $c_s$, we obtain
\be
 \Delta k =  \beta k_0   \ll k_0 \;. \label{DeltaK}
\ee

Plugging the expressions of $k_0$ and $\Delta k$ into (\ref{Occupation}) and using $n_\gamma \simeq \omega (\MP h^+_0)^2$, the occupation number can be written as
  \begin{equation}
 n^\pi_{k} =  {4 \beta c_s^7 \pi^2 }   \left( \frac{\Lambda}{\omega} \right)^4\frac{n_\pi}{n_\gamma}
  \ , \label{Occupation1}
 \end{equation}
where we have  focused on the case of the operator $m_3^3$ (which can be straightforwardly extended to the case of the operator $\tilde m_4^2$ by the replacement $\Lambda^2 \to \Lambda_\star^3 \omega^{-1}$, see discussion in Sec.~\ref{tildem42op}) and used the definition of $\beta$, eq.~\eqref{betadef1}, to rewrite $(\MP h^+_0)^2$. The Bose condensation effect becomes important for $n_k^\pi \gg 1$ or, using the above equation, for 
 \begin{equation}
 n_\pi \gg \frac{n_\gamma}{ 4 \beta c_s^7 \pi^2 } \left( \frac{\omega}{\Lambda} \right)^4  \ . \label{BoseCondition}
 \end{equation}
In this case, we can solve eq.~\eqref{RateOfChange} with the decay rate given by eq.~\eqref{DecatRate}. This gives
\be
n_\pi \propto \exp \left( \frac{\pi \beta  }{30} \omega u \right) \;,
\ee
which displays an instability similar to that encountered above in eq.~\eqref{eq:pienergy}, but with a different exponent. Notice that the approach of Appendix~\ref{Bose} is approximate and does not reproduce the correct numerical factors in the timescale of the instability. Of course, a more precise calculation would give the same answer.

 It is useful to check that our formula for the modification of the GW, eq.~\eqref{eq:GammaBR}, smoothly interpolates with the perturbative decay result, eq.~\eqref{DecatRate}, when the occupation number becomes small. In the regime in which Bose enhancement is negligible $n_\pi/n_\gamma \sim \Gamma u$, see Fig.~\ref{fig:diagram}.  Using this in eq.~\eqref{Occupation1} we get  $n^\pi_{k} \sim \beta \omega u$. Not surprisingly this is the parameter that enters the exponential growth of the instability. Our saddle-point treatment is valid for  
$\beta \omega u \gg 1$, but we expect that, when $\beta \omega u \sim 1$, it gives a result of the same order as the perturbative decay of gravitons. Indeed if we plug this equality in eq.~\eqref{eq:GammaBR} one gets (for $c_s \ll 1$)
\be
\Delta \gamma \simeq \frac{v}{\Lambda^2} \frac{1}{c_s^5}\beta \omega^4 \simeq \MP h_0^+ \Gamma v \;.
\ee
This is indeed the perturbative result: the original GW, $\MP h_0^+$, changes with a rate $\Gamma$ for a time of order $v$.

\section{Details on the conservation of energy\label{app:energy}}

\subsection{$m_3^3$-operator}
In this Appendix we check the conservation of energy discussed in Section \ref{sec:conservation}.
First of all, let us verify that GWs do not contribute to the flux of energy across $\partial\mathcal V_0$ (see Fig.~\ref{fig:diagram_boundary}). This is clearly true for $\bar\gamma_{ij}$ but it holds at order $\Delta\gamma$ too. Indeed we have, using $\dot{\bar\gamma}_{ij} = - \partial_z \bar\gamma_{ij}$ and $\partial_k \bar\gamma_{ij} = \delta_{k z} \partial_z \bar\gamma_{ij}$,
\begin{align}\label{T00-T0z}
\left(T^{00} - T^{z0}\right)\big\rvert_{u = |z_0|} &\supset \frac{1}{4}\left[ (\dot{\bar\gamma}_{ij})^2 +( \partial_k \bar\gamma_{ij})^2 \right]  + \frac{1}{2}\dot{\bar \gamma}_{ij} \partial_z \bar\gamma_{ij}  \\
&  \nonumber + \frac{1}{2}\dot{\bar\gamma}_{ij} \Delta\dot \gamma_{ij} + \frac{1}{2}\partial_k \bar\gamma_{ij} \partial_k \Delta\gamma_{ij} + \frac{1}{2} \dot{\bar \gamma}_{ij}\partial_z \Delta\gamma_{ij}  +\frac{1}{2} \Delta\dot \gamma_{ij}\partial_z\bar\gamma_{ij} \\
& \nonumber =  0  + \frac{1}{2}\dot {\bar\gamma}_{ij} ( \Delta\dot \gamma_{ij} - \partial_z \Delta\gamma_{ij} ) + \frac{1}{2} \dot {\bar\gamma}_{ij} (\partial_z \Delta\gamma_{ij} - \Delta\dot \gamma_{ij}  ) = 0\;.
\end{align}

Let us now calculate the LHS of eq.~\eqref{eq:energy_conservation},  {\em i.e.}~the variation of the total energy in the region. This is only due to $\Delta\gamma_{ij}$, since $\pi$ and $\bar\gamma_{ij}$ depend on $u$ only. One has
\begin{align}
T^{00} &\supset \frac{1}{2} \dot {\bar\gamma}_{ij} \Delta\dot \gamma_{ij} + \frac{1}{2}\partial_k \bar\gamma_{ij} \partial_k \Delta\gamma_{ij} = \frac{1}{2} \dot{\bar\gamma}_{ij} \left( \Delta \dot \gamma_{ij} - \partial_z \Delta\gamma_{ij}\right) = \frac{1}{2} \dot{\bar\gamma}_{ij} 2 \partial_u \Delta\gamma_{ij}  \\
& \nonumber = -\dot{\bar\gamma}_{ij} \left( \frac{v}{4 \Lambda^2}\partial_u J_{ij} (u)\right) = -\dot{\bar\gamma}_{ij} \epsilon^+_{ij} \left( \frac{v}{4 \Lambda^2}\partial_u \braket{(\partial_x \pi)^2 -(\partial_y \pi)^2}\right) \equiv v \mathcal F (u)\;.
\end{align}
The integral over $\partial \mathcal V _2 \cup \partial \mathcal V _1$ reduces to 
\begin{align}\label{integral_T00}
& \int_{\partial \mathcal V _2 } T^{00}\, \text d z - \int_{\partial \mathcal V _1 } T^{00}\, \text d z = \int_{T-|z_0|}^T v \mathcal F(u)\big\rvert_{t = T} \, \text d z - \int_{-|z_0|}^0 v \mathcal F(u)\big\rvert_{t = 0} \, \text d z = \\
& \nonumber \int_{T-|z_0|}^T (T  + z) \mathcal F(T-z)\, \text d z - \int_{-|z_0|}^0 z \mathcal F(-z)\, \text d z
= \int_{-|z_0|}^0 (2 T + \tilde z) \mathcal F(-\tilde z)\, \text d \tilde z -\int_{-|z_0|}^0 z \mathcal F(-z)\, \text d z \\
& \nonumber =  2 T \int_{-|z_0|}^0  \mathcal F(-z)\, \text d z \, .
\end{align}
The RHS of eq.~\eqref{eq:energy_conservation} gets contribution only from $\pi$: 
\be
\int_{\partial \mathcal V_0}  \left( T^{00} - T^{z0}\right)\text d t  = \int_0^T \left(T^{00}_\pi\ - T^{z0}_\pi \right)\big\rvert_{u = |z_0|}  \text d t 
= T \left(T^{00}_\pi - T^{z0}_\pi \right)\big\rvert_{u = |z_0|} \;.
\ee
The equation for energy conservation, eq.~\eqref{eq:energy_conservation}, becomes
\begin{align}
0 &= T \left(T^{00}_\pi - T^{z0}_\pi \right)\big\rvert_{u = |z_0|}  + 2 T \int_{-|z_0|}^0  \mathcal F(-z)\, \text d z \\
& = T \left[ \left(T^{00}_\pi - T^{z0}_\pi \right)\big\rvert_{u = |z_0|} -  \frac{1}{2 \Lambda^2} \int_{0}^{|z_0|}\dot{\bar\gamma}_{ij}(u) \epsilon^+_{ij}\partial_u \braket{(\partial_x \pi)^2 - (\partial_y \pi)^2} \, \text d u\right]\;.\label{eq:energy_continued}
\end{align}
Note that the linear $v$ dependence of $\Delta\gamma$ is essential: if it was not the case then the two terms in \eqref{eq:energy_continued} would have a different $T$ dependence, with no chance of adding up to zero. To explicitly check energy conservation one should integrate \eqref{eq:energy_continued}. A faster way to check this is to take a derivative with respect to $|z_0|$ (or equivalently $u$). The resulting equation can be shown to be satisfied by the solution for  $\pi$, \eqref{eq:3}. After simplifying $T$ and taking the derivative with respect to $z_0$, \eqref{eq:energy_continued} becomes
\begin{align}
\partial_{u} \left[ \braket{T^{00}_\pi( u)} - \braket{T^{z0}_\pi ( u)} \right] - \frac{1}{2 \Lambda^2}  \dot{\bar\gamma}_{ij}(u) \epsilon^+_{ij}\partial_{u} \braket{(\partial_x \pi)^2 - (\partial_y \pi)^2} = 0\;.
\end{align}
The expression of $\langle T^{00}_\pi \rangle$ is given by eq.~\eqref{eq:rhopi} while
\be
\braket{T^{z0}_\pi (u)}  =-c_s^2\braket{\dot \pi \partial_z \pi} 
= -\int \frac{\text d ^3 \tilde{\vect p}}{(2 \pi)^3}\frac{1}{ 4 c_s^2 p_u}\left[ -2 c_s^2 |\partial_uf_{\tilde{p}}|^2  -2 p_s^2 |f_{\tilde{p}}|^2 + {\rm const}   \right]\;.
\ee
Therefore we can write
\begin{equation}
\partial_{u} \left[ \braket{T^{00}_\pi( u)} - \braket{T^{z0}_\pi ( u)} \right] = \int \frac{\text d ^3 \tilde{\vect p}}{(2 \pi)^3}\frac{1}{ 4 c_s^2 p_u}\left[{\partial_u f_{\tilde{p}}^\star} \left((1-c_s^2) \partial_u^2 f_{\tilde{p}} + f_{\tilde{p}} \left( (1-c_s^2)c_s^{-2} p_s^2 + c_s^2 (p_x^2  + p_y^2)\right) \right) +  {\rm h.c.}\right].
\end{equation}
The remaining term contains
\be
\partial_u\braket{(\partial_x \pi)^2 - (\partial_y \pi)^2}  = \partial_u \int \frac{\text d ^3 \tilde{\vect p}}{(2 \pi)^3}\frac{1}{ 4 c_s^2 p_u} 2 (p_x^2 - p_y^2)|f_{\tilde{p}}|^2 
=  \int \frac{\text d ^3 \tilde{\vect p}}{(2 \pi)^3}\frac{1}{ 4 c_s^2 p_u} 2 (p_x^2 - p_y^2){f_{\tilde{p}}^{\star}}'f_{\tilde{p}} + {\rm h.c.}
\ee
At this point by adding up these contributions we can collect the terms with ${\partial_u f_{\tilde{p}}^{\star}}$. They are
\begin{equation}
{\partial_u f_{\tilde{p}}^{\star}} \left[ (1-c_s^2)\partial_u^2 f_{\tilde{p}} + f_{\tilde{p}} \left( (1-c_s^2)c_s^{-2} p_s^2 + c_s^2 (p_x^2  + p_y^2) - \frac{1}{ \Lambda^2}\dot\gamma^b_{ij}(u) \epsilon^+_{ij} (p_x^2 - p_y^2) \right)\right] + {\rm h.c.}
\end{equation}
This equation is solved by $f_{\tilde{p}}$: this is easily seen by comparing with equation \eqref{eq:3}.

\subsection{\texorpdfstring{$\tilde m_4^2$}{tilde m42}-operator}
In this section we are going to use the same logic as the one in the previous section to verify the conservation of energy for $\tilde{m}_4^2$-operator. According to the Lagrangian (\ref{ddotgamma}), the components of the energy-momentum tensor are given by
\begin{align}
T^0_{\ 0} &=  -\left[\frac{1}{4} \dot{\gamma}^2_{ij} + \frac{1}{4}(\partial_k\gamma_{ij})^2 + \frac{1}{2}\dot{\pi}^2 + \frac{1}{2}c_s^2(\partial_i\pi)^2 - \frac{2}{\Lambda_\star^3}\dot{\gamma}_{ij}\partial_i\dot{\pi}\partial_j\pi \right]  \; \label{T_00_m4tilde}, \\
T^0_{\ i} &= -\frac{1}{2}\dot{\gamma}_{kl}\partial_i\gamma_{kl} - \dot{\pi}\partial_i\pi + \frac{2}{\Lambda_\star^3}\partial_i\gamma_{kl}\partial_k\dot{\pi}\partial_l\pi - \frac{1}{\Lambda_\star^3}\partial_i\dot{\gamma}_{kl}\partial_k\pi\partial_l\pi \; , \label{T_0i_m4tilde} \\
T^i_{\ 0} &= \frac{1}{2}\dot{\gamma}_{kl}\partial_i\gamma_{kl} + c_s^2\dot{\pi}\partial_i\pi - \frac{2}{\Lambda_\star^3}\ddot{\gamma}_{ij}\dot{\pi}\partial_j\pi \;   , \\
T^i_{\ j} &= \frac{1}{2}\partial_i\gamma_{kl}\partial_j\gamma_{kl} + c_s^2\partial_i\pi\partial_j\pi  - \frac{2}{\Lambda_\star^3}\ddot\gamma_{ik}\partial_j\pi\partial_k\pi \nonumber \\  & \quad + \frac{1}{2}\delta^i_j\left[\frac{1}{2}\dot{\gamma}_{kl}^2 - \frac{1}{2}(\partial_m\gamma_{kl})^2 + \dot{\pi}^2 -c_s^2(\partial_l\pi)^2 + \frac{2}{\Lambda_\star^3}\ddot{\gamma}_{kl}\partial_k\pi\partial_l\pi \right] \; \label{m4_tildeT_ij}. 
\end{align}
Notice that there is an extra term in $T^0_{\ 0}$ due to the interaction $\gamma\pi\pi$, unlike the case of $m_3^3$-operator. Since this new piece is second order in $\pi$, it can be approximated as $- \frac{2}{\Lambda_\star^3}\dot{\bar{\gamma}}_{ij}\partial_i\dot{\pi}\partial_j\pi$. 

Let us first consider the RHS of (\ref{eq:energy_conservation}), taking into account that $\dot{\bar\gamma}_{ij} = - \partial_z \bar\gamma_{ij}$ and $\partial_k \bar\gamma_{ij} = \delta_{k z} \partial_z \bar\gamma_{ij}$,
\begin{align}\label{T00-T0z_m4tilde}
\int_{\partial \mathcal V_0} \left(T^{00} - T^{z0}\right)\big\rvert_{u = |z_0|} &=  T\left(\frac{1}{2}\dot{\pi}^2 + \frac{c_s^2}{2}(\partial_k\pi)^2 + c_s^2\dot{\pi}\partial_z\pi - \frac{2}{\Lambda_\star^3}\dot{\bar{\gamma}}_{ij}\partial_i\dot{\pi}\partial_j\pi \right)_{u = |z_0|}\nonumber \\
&\equiv T\left(T^{00}_\pi - T^{z0}_\pi - \frac{2}{\Lambda_\star^3}\dot{\bar{\gamma}}_{ij}\partial_i\dot{\pi}\partial_j\pi \right)_{u = |z_0|} \; ,
\end{align}
where all terms involving only GWs  added up to zero because of eq.~(\ref{T00-T0z}).
The LHS of (\ref{eq:energy_conservation}) gets contributions only from $\Delta\gamma_{ij}$, as for the ${m}_3^3$-operator case. We have
\begin{align}
T^{00} 
& \supset \dot{\bar\gamma}_{ij} \left( \frac{v}{4 \Lambda_\star^3}\partial_u^2 J_{ij} (u)\right) = \dot{\bar\gamma}_{ij} \epsilon^+_{ij} \left( \frac{v}{4 \Lambda_\star^3}\partial_u^2 \braket{(\partial_x \pi)^2 -(\partial_y \pi)^2}\right) \equiv v \tilde{\mathcal{F}} (u)\;.
\end{align}
Integrating $T^{00}$ over $\partial \mathcal V _2 \cup \partial \mathcal V _1$ gives
\begin{align}\label{Integral_T00_m4tilde}
& \int_{\partial \mathcal V _2 } T^{00}\, \text d z - \int_{\partial \mathcal V _1 } T^{00}\, \text d z = 2 T \int_{-|z_0|}^0  \tilde{\mathcal{F}}(-z)\, \text d z \, ,
\end{align}
similarly to eq.~(\ref{integral_T00}).
Using (\ref{T00-T0z_m4tilde}) and (\ref{Integral_T00_m4tilde}), eq.~(\ref{eq:energy_conservation}) becomes
\begin{align}
0 &= T \left(T^{00} - T^{z0} \right)\big\rvert_{u = |z_0|}  + 2 T \int_{-|z_0|}^0  \tilde{\mathcal{F}}(-z)\, \text d z \nonumber \\
& = T \left[ \left(T^{00}_\pi - T^{z0}_\pi - \frac{2}{\Lambda_\star^3}\dot{\bar{\gamma}}_{ij}\partial_i\dot{\pi}\partial_j\pi \right)_{u = |z_0|} +  \frac{1}{2 \Lambda^3_\star} \int_{0}^{|z_0|}\dot{\bar\gamma}_{ij}(u) \epsilon^+_{ij}\partial^2_u \braket{(\partial_x \pi)^2 - (\partial_y \pi)^2} \, \text d u\right]\;.\label{eq:energy_continued_m4 tilde}
\end{align}
Like in the previous section, one can verify this equation by taking a derivative with respect to $|z_0|$ (or equivalently $u$): 
\begin{align}
\partial_{u} \left[ \braket{T^{00}_\pi( u)} - \braket{T^{z0}_\pi ( u)} \right] - \frac{1}{ \Lambda_\star^3}  \ddot{\bar\gamma}_{ij}(u) \epsilon^+_{ij}\partial_{u} \braket{(\partial_x \pi)^2 - (\partial_y \pi)^2} = 0\;.
\end{align}
As we have shown before, the first term of LHS can be expressed as
\begin{align}\label{T00-T0z_m4tilde2}
\partial_{u} \left[ \braket{T^{00}_\pi( u)} - \braket{T^{z0}_\pi ( u)} \right] = \int \frac{\text d ^3 \tilde{\vect p}}{(2 \pi)^3}\frac{1}{ 4 c_s^2 p_u}\left[{\partial_u f_{\tilde{p}}^\star} \left((1-c_s^2) \partial_u^2 f_{\tilde{p}} + f_{\tilde{p}} \left( (1-c_s^2)c_s^{-2} p_s^2 + c_s^2 (p_x^2  + p_y^2)\right) \right) +  {\rm h.c.}\right].
\end{align}
Similarly, the second term can be rewritten as 
\begin{align}\label{interaction_m4tilde}
- \frac{1}{ \Lambda_\star^3} \ddot{\bar\gamma}_{ij}(u) \epsilon^+_{ij}\partial_u\braket{(\partial_x \pi)^2 - (\partial_y \pi)^2}  =   - \frac{1}{ \Lambda_\star^3} \ddot{\bar\gamma}_{ij}(u) \epsilon^+_{ij}\int \frac{\text d ^3 \tilde{\vect p}}{(2 \pi)^3}\frac{1}{ 4 c_s^2 p_u} 2 (p_x^2 - p_y^2){f_{\tilde{p}}^{\star}}'f_{\tilde{p}} + {\rm h.c.}
\end{align}
Adding up (\ref{T00-T0z_m4tilde2}) and (\ref{interaction_m4tilde}) together one therefore obtains
\begin{align}
{\partial_u f_{\tilde{p}}^{\star}} \left[ (1-c_s^2)\partial_u^2 f_{\tilde{p}} + f_{\tilde{p}} \left( (1-c_s^2)c_s^{-2} p_s^2 + c_s^2 (p_x^2  + p_y^2) - \frac{1}{ \Lambda_\star^3}\ddot\gamma^b_{ij}(u) \epsilon^+_{ij} (p_x^2 - p_y^2) \right)\right] + {\rm h.c.}
\end{align}
This coincides with the equation of motion for $f_{\tilde{p}}$, which can be obtained by expanding eq.~(\ref{EoM3}) in Fourier modes.


\small{
\bibliography{bib_v3}}

\providecommand{\href}[2]{#2}\begingroup\raggedright\begin{thebibliography}{10}

\bibitem{Creminelli:2018xsv}
P.~Creminelli, M.~Lewandowski, G.~Tambalo, and F.~Vernizzi, ``{Gravitational
  Wave Decay into Dark Energy},'' {\em JCAP} {\bf 1812} (2018), no.~12 025,
  \href{http://xxx.lanl.gov/abs/1809.03484}{{\tt 1809.03484}}.

\bibitem{Creminelli:2006xe}
P.~Creminelli, M.~A. Luty, A.~Nicolis, and L.~Senatore, ``{Starting the
  Universe: Stable Violation of the Null Energy Condition and Non-standard
  Cosmologies},'' {\em JHEP} {\bf 0612} (2006) 080,
  \href{http://xxx.lanl.gov/abs/hep-th/0606090}{{\tt hep-th/0606090}}.

\bibitem{Cheung:2007st}
C.~Cheung, P.~Creminelli, A.~L. Fitzpatrick, J.~Kaplan, and L.~Senatore, ``{The
  Effective Field Theory of Inflation},'' {\em JHEP} {\bf 0803} (2008) 014,
  \href{http://xxx.lanl.gov/abs/0709.0293}{{\tt 0709.0293}}.

\bibitem{Creminelli:2008wc}
P.~Creminelli, G.~D'Amico, J.~Norena, and F.~Vernizzi, ``{The Effective Theory
  of Quintessence: the $w<-1$ Side Unveiled},'' {\em JCAP} {\bf 0902} (2009)
  018, \href{http://xxx.lanl.gov/abs/0811.0827}{{\tt 0811.0827}}.

\bibitem{Gubitosi:2012hu}
G.~Gubitosi, F.~Piazza, and F.~Vernizzi, ``{The Effective Field Theory of Dark
  Energy},'' {\em JCAP} {\bf 1302} (2013) 032,
  \href{http://xxx.lanl.gov/abs/1210.0201}{{\tt 1210.0201}}.

\bibitem{Gleyzes:2013ooa}
J.~Gleyzes, D.~Langlois, F.~Piazza, and F.~Vernizzi, ``{Essential Building
  Blocks of Dark Energy},'' {\em JCAP} {\bf 1308} (2013) 025,
  \href{http://xxx.lanl.gov/abs/1304.4840}{{\tt 1304.4840}}.

\bibitem{TheLIGOScientific:2017qsa}
{\bf Virgo, LIGO Scientific} Collaboration, B.~P. Abbott {\em et.~al.},
  ``{GW170817: Observation of Gravitational Waves from a Binary Neutron Star
  Inspiral},'' {\em Phys. Rev. Lett.} {\bf 119} (2017), no.~16 161101,
  \href{http://xxx.lanl.gov/abs/1710.05832}{{\tt 1710.05832}}.

\bibitem{Creminelli:2017sry}
P.~Creminelli and F.~Vernizzi, ``{Dark Energy after GW170817 and GRB170817A},''
  {\em Phys. Rev. Lett.} {\bf 119} (2017), no.~25 251302,
  \href{http://xxx.lanl.gov/abs/1710.05877}{{\tt 1710.05877}}.

\bibitem{Sakstein:2017xjx}
J.~Sakstein and B.~Jain, ``{Implications of the Neutron Star Merger GW170817
  for Cosmological Scalar-Tensor Theories},'' {\em Phys. Rev. Lett.} {\bf 119}
  (2017), no.~25 251303, \href{http://xxx.lanl.gov/abs/1710.05893}{{\tt
  1710.05893}}.

\bibitem{Ezquiaga:2017ekz}
J.~M. Ezquiaga and M.~Zumalacrregui, ``{Dark Energy After GW170817: Dead Ends
  and the Road Ahead},'' {\em Phys. Rev. Lett.} {\bf 119} (2017), no.~25
  251304, \href{http://xxx.lanl.gov/abs/1710.05901}{{\tt 1710.05901}}.

\bibitem{Baker:2017hug}
T.~Baker, E.~Bellini, P.~G. Ferreira, M.~Lagos, J.~Noller, and I.~Sawicki,
  ``{Strong constraints on cosmological gravity from GW170817 and GRB
  170817A},'' {\em Phys. Rev. Lett.} {\bf 119} (2017), no.~25 251301,
  \href{http://xxx.lanl.gov/abs/1710.06394}{{\tt 1710.06394}}.

\bibitem{deRham:2018red}
C.~de~Rham and S.~Melville, ``{Gravitational Rainbows: LIGO and Dark Energy at
  its Cutoff},'' {\em Phys. Rev. Lett.} {\bf 121} (2018), no.~22 221101,
  \href{http://xxx.lanl.gov/abs/1806.09417}{{\tt 1806.09417}}.

\bibitem{Creminelli:2019new}
P.~Creminelli, G.~Tambalo, F.~Vernizzi, and V.~Yingcharoenrat, ``{Dark Energy
  Instabilities induced by Gravitational Waves},'' {\em in preparation}.

\bibitem{Caldwell:1997ii}
R.~R. Caldwell, R.~Dave, and P.~J. Steinhardt, ``{Cosmological imprint of an
  energy component with general equation of state},'' {\em Phys. Rev. Lett.}
  {\bf 80} (1998) 1582--1585,
  \href{http://xxx.lanl.gov/abs/astro-ph/9708069}{{\tt astro-ph/9708069}}.

\bibitem{ArmendarizPicon:2000dh}
C.~Armendariz-Picon, V.~F. Mukhanov, and P.~J. Steinhardt, ``{A Dynamical
  solution to the problem of a small cosmological constant and late time cosmic
  acceleration},'' {\em Phys.Rev.Lett.} {\bf 85} (2000) 4438--4441,
  \href{http://xxx.lanl.gov/abs/astro-ph/0004134}{{\tt astro-ph/0004134}}.

\bibitem{Deffayet:2010qz}
C.~Deffayet, O.~Pujolas, I.~Sawicki, and A.~Vikman, ``{Imperfect Dark Energy
  from Kinetic Gravity Braiding},'' {\em JCAP} {\bf 1010} (2010) 026,
  \href{http://xxx.lanl.gov/abs/1008.0048}{{\tt 1008.0048}}.

\bibitem{Kobayashi:2010cm}
T.~Kobayashi, M.~Yamaguchi, and J.~Yokoyama, ``{G-inflation: Inflation driven
  by the Galileon field},'' {\em Phys. Rev. Lett.} {\bf 105} (2010) 231302,
  \href{http://xxx.lanl.gov/abs/1008.0603}{{\tt 1008.0603}}.

\bibitem{Horndeski:1974wa}
G.~W. Horndeski, ``{Second-order scalar-tensor field equations in a
  four-dimensional space},'' {\em Int.J.Theor.Phys.} {\bf 10} (1974) 363--384.

\bibitem{Deffayet:2011gz}
C.~Deffayet, X.~Gao, D.~Steer, and G.~Zahariade, ``{From k-essence to
  generalised Galileons},'' {\em Phys.Rev.} {\bf D84} (2011) 064039,
  \href{http://xxx.lanl.gov/abs/1103.3260}{{\tt 1103.3260}}.

\bibitem{Gleyzes:2014dya}
J.~Gleyzes, D.~Langlois, F.~Piazza, and F.~Vernizzi, ``{Healthy theories beyond
  Horndeski},'' {\em Phys. Rev. Lett.} {\bf 114} (2015), no.~21 211101,
  \href{http://xxx.lanl.gov/abs/1404.6495}{{\tt 1404.6495}}.

\bibitem{DAmico:2016ntq}
G.~D'Amico, Z.~Huang, M.~Mancarella, and F.~Vernizzi, ``{Weakening Gravity on
  Redshift-Survey Scales with Kinetic Matter Mixing},'' {\em JCAP} {\bf 1702}
  (2017) 014, \href{http://xxx.lanl.gov/abs/1609.01272}{{\tt 1609.01272}}.

\bibitem{Gleyzes:2014qga}
J.~Gleyzes, D.~Langlois, F.~Piazza, and F.~Vernizzi, ``{Exploring gravitational
  theories beyond Horndeski},'' {\em JCAP} {\bf 1502} (2015) 018,
  \href{http://xxx.lanl.gov/abs/1408.1952}{{\tt 1408.1952}}.

\bibitem{Cusin:2017mzw}
G.~Cusin, M.~Lewandowski, and F.~Vernizzi, ``{Nonlinear Effective Theory of
  Dark Energy},'' {\em JCAP} {\bf 1804} (2018), no.~04 061,
  \href{http://xxx.lanl.gov/abs/1712.02782}{{\tt 1712.02782}}.

\bibitem{Bellini:2014fua}
E.~Bellini and I.~Sawicki, ``{Maximal freedom at minimum cost: linear
  large-scale structure in general modifications of gravity},'' {\em JCAP} {\bf
  1407} (2014) 050, \href{http://xxx.lanl.gov/abs/1404.3713}{{\tt 1404.3713}}.

\bibitem{Gleyzes:2014rba}
J.~Gleyzes, D.~Langlois, and F.~Vernizzi, ``{A unifying description of dark
  energy},'' {\em Int. J. Mod. Phys.} {\bf D23} (2015), no.~13 1443010,
  \href{http://xxx.lanl.gov/abs/1411.3712}{{\tt 1411.3712}}.

\bibitem{Langlois:2015cwa}
D.~Langlois and K.~Noui, ``{Degenerate higher derivative theories beyond
  Horndeski: evading the Ostrogradski instability},'' {\em JCAP} {\bf 1602}
  (2016), no.~02 034, \href{http://xxx.lanl.gov/abs/1510.06930}{{\tt
  1510.06930}}.

\bibitem{Crisostomi:2016czh}
M.~Crisostomi, K.~Koyama, and G.~Tasinato, ``{Extended Scalar-Tensor Theories
  of Gravity},'' {\em JCAP} {\bf 1604} (2016), no.~04 044,
  \href{http://xxx.lanl.gov/abs/1602.03119}{{\tt 1602.03119}}.

\bibitem{Langlois:2017mxy}
D.~Langlois, M.~Mancarella, K.~Noui, and F.~Vernizzi, ``{Effective Description
  of Higher-Order Scalar-Tensor Theories},'' {\em JCAP} {\bf 1705} (2017),
  no.~05 033, \href{http://xxx.lanl.gov/abs/1703.03797}{{\tt 1703.03797}}.

\bibitem{mclachlan1951theory}
N.~W. McLachlan, {\em Theory and application of Mathieu functions}.
\newblock Clarendon Press, 1951.

\bibitem{olver2010nist}
F.~W. Olver, D.~W. Lozier, R.~F. Boisvert, and C.~W. Clark, {\em NIST handbook
  of mathematical functions hardback and CD-ROM}.
\newblock Cambridge university press, 2010.

\bibitem{Maggiore:1900zz}
M.~Maggiore, {\em {Gravitational Waves. Vol. 1: Theory and Experiments}}.
\newblock Oxford Master Series in Physics. Oxford University Press, 2007.

\bibitem{Wald:1984rg}
R.~M. Wald, {\em {General Relativity}}.
\newblock Chicago Univ. Pr., Chicago, USA, 1984.

\bibitem{Prokopec:1996rr}
T.~Prokopec and T.~G. Roos, ``{Lattice study of classical inflaton decay},''
  {\em Phys. Rev.} {\bf D55} (1997) 3768--3775,
  \href{http://xxx.lanl.gov/abs/hep-ph/9610400}{{\tt hep-ph/9610400}}.

\bibitem{Adshead:2017xll}
P.~Adshead, J.~T. Giblin, and Z.~J. Weiner, ``{Non-Abelian gauge preheating},''
  {\em Phys. Rev.} {\bf D96} (2017), no.~12 123512,
  \href{http://xxx.lanl.gov/abs/1708.02944}{{\tt 1708.02944}}.

\bibitem{Pirtskhalava:2015nla}
D.~Pirtskhalava, L.~Santoni, E.~Trincherini, and F.~Vernizzi, ``{Weakly Broken
  Galileon Symmetry},'' {\em JCAP} {\bf 1509} (2015), no.~09 007,
  \href{http://xxx.lanl.gov/abs/1505.00007}{{\tt 1505.00007}}.

\bibitem{Santoni:2018rrx}
L.~Santoni, E.~Trincherini, and L.~G. Trombetta, ``{Behind Horndeski:
  Structurally Robust Higher Derivative EFTs},'' {\em JHEP} {\bf 08} (2018)
  118, \href{http://xxx.lanl.gov/abs/1806.10073}{{\tt 1806.10073}}.

\bibitem{Crisostomi:2019yfo}
M.~Crisostomi, M.~Lewandowski, and F.~Vernizzi, ``{Vainshtein regime in
  Scalar-Tensor gravity: constraints on DHOST theories},''
  \href{http://xxx.lanl.gov/abs/1903.11591}{{\tt 1903.11591}}.

\bibitem{Audley:2017drz}
{\bf LISA} Collaboration, H.~Audley {\em et.~al.}, ``{Laser Interferometer
  Space Antenna},'' \href{http://xxx.lanl.gov/abs/1702.00786}{{\tt
  1702.00786}}.

\bibitem{Abbott:2016blz}
{\bf Virgo, LIGO Scientific} Collaboration, B.~P. Abbott {\em et.~al.},
  ``{Observation of Gravitational Waves from a Binary Black Hole Merger},''
  {\em Phys. Rev. Lett.} {\bf 116} (2016), no.~6 061102,
  \href{http://xxx.lanl.gov/abs/1602.03837}{{\tt 1602.03837}}.

\bibitem{Malik:2008im}
K.~A. Malik and D.~Wands, ``{Cosmological perturbations},'' {\em Phys. Rept.}
  {\bf 475} (2009) 1--51, \href{http://xxx.lanl.gov/abs/0809.4944}{{\tt
  0809.4944}}.

\bibitem{Maldacena:2002vr}
J.~M. Maldacena, ``{Non-Gaussian features of primordial fluctuations in single
  field inflationary models},'' {\em JHEP} {\bf 0305} (2003) 013,
  \href{http://xxx.lanl.gov/abs/astro-ph/0210603}{{\tt astro-ph/0210603}}.

\bibitem{Mukhanov:2005sc}
V.~Mukhanov, {\em {Physical Foundations of Cosmology}}.
\newblock Cambridge University Press, Oxford, 2005.

\end{thebibliography}\endgroup
\bibliographystyle{utphys}

\end{document}